\documentclass[usenatbib,a4paper,times,fleqn]{mnras}
\usepackage{graphicx,natbib,times,amssymb,amsmath,pstricks,mathtools,enumitem,array}
\usepackage{aas_macros}
\usepackage{bm,url}
\usepackage{breakurl}

\newcommand {\apgt} {\ {\raise-.5ex\hbox{$\buildrel>\over\sim$}}\ }
\newcommand {\aplt} {\ {\raise-.5ex\hbox{$\buildrel<\over\sim$}}\ }

\title[Growth with Discontinuous Galerkin schemes]{Grain growth for astrophysics with Discontinuous Galerkin schemes}

\author[Lombart \& Laibe]{Maxime Lombart$^{1}$\thanks{maxime.lombart@ens-lyon.fr}, Guillaume Laibe$^{1,2}$\thanks{guillaume.laibe@ens-lyon.fr} \\
$^{1}$Univ Lyon, Univ Lyon1, Ens de Lyon, CNRS, Centre de Recherche Astrophysique de Lyon UMR5574, F-69230, Saint-Genis,-Laval, France.\\
$^{2}$Institut Universitaire de France\\
}
\pagerange{\pageref{firstpage}--\pageref{lastpage}} \pubyear{2020}
\date{Accepted 2020 November 23. Received 2020 November 19; in original form 2020 October 20.}
\begin{document}
\label{firstpage}

\maketitle

\begin{abstract}
Depending on their sizes, dust grains store more or less charges, catalyse more or less chemical reactions, intercept more or less photons and stick more or less efficiently to form embryos of planets. Hence the need for an accurate treatment of dust coagulation and fragmentation in numerical modelling. However, existing algorithms for solving the coagulation equation are over-diffusive in the conditions of 3D simulations. We address this challenge by developing a high-order solver based on the Discontinuous Galerkin method. This algorithm conserves mass to machine precision and allows to compute accurately the growth of dust grains over several orders of magnitude in size with a very limited number of dust bins.

\end{abstract}

\section{Introduction}
\label{sec:introduction}
Solid particles pervade the interstellar medium at all scales. Although they represent a small amount of its total mass, they deeply influence its evolution by setting the local chemical, thermal and charge balances. Dust plays also a key role in the formation of planets, since solid bodies grow over thirty orders of magnitude in mass to form cores of planets. Spatially resolved observations of young stellar objects strongly suggest that at least some planets have to form in less that one million of years (e.g. \citealt{Alma2015,Avenhaus2018,Pinte2020}). Key is to understand how dust growth can be so efficient. However, planet formation is an out-of-equilibrium non-linear multi-scales and multi-physics process. For example, dust grains differentiate from the gas as they settle vertically and drift radially in the disc (i.e \citealt{Testi2014} and references therein). This creates instabilities which concentrate the solids even more, affecting the collisional rate of the grains, and thus, their growth or fragmentation. Since dust dynamics strongly depends on the grain size, growth operates a strong feed-back on the spatial distribution of the particles.

\begin{figure}
\includegraphics[width=0.95\columnwidth]{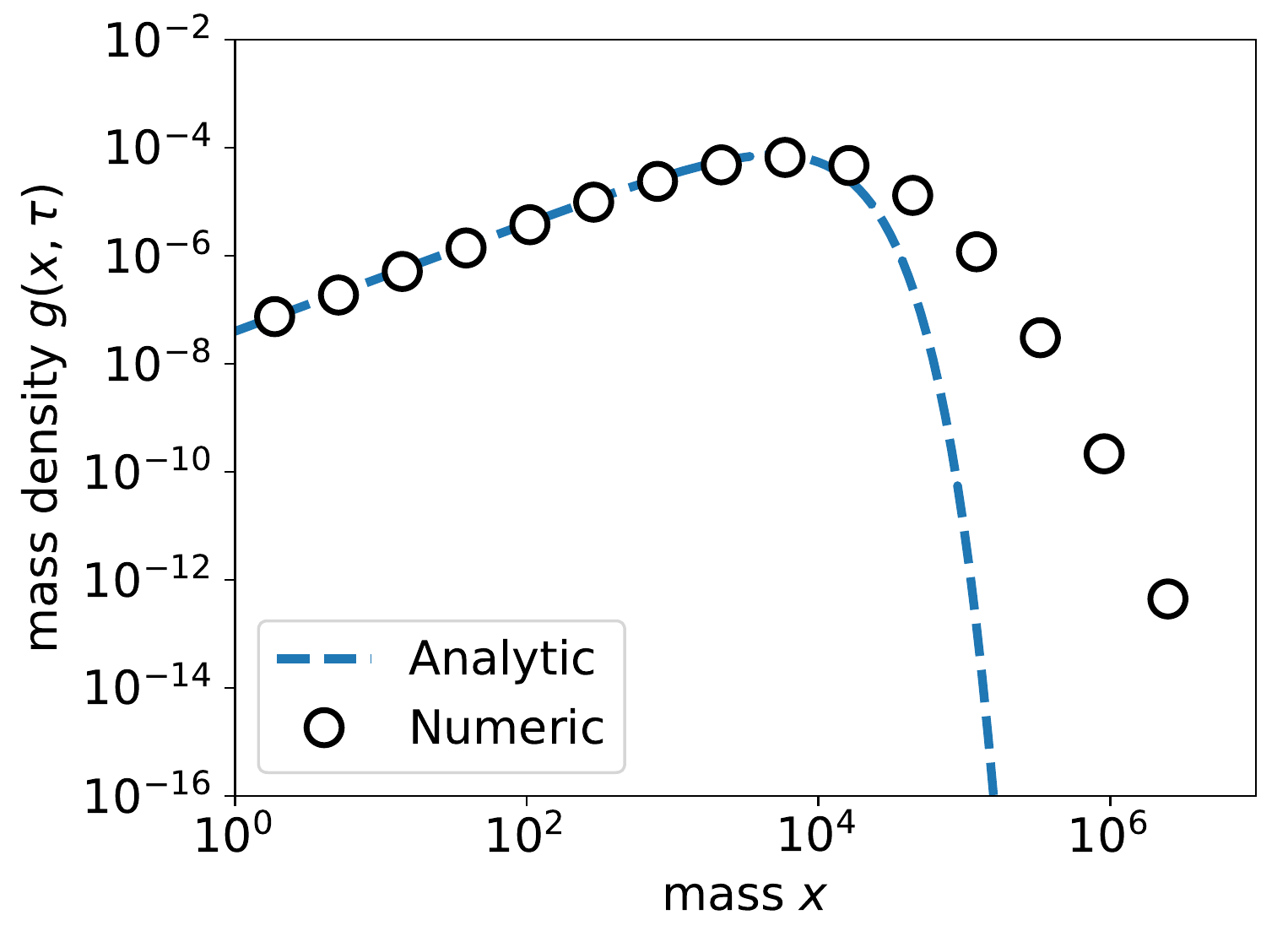}
\caption{An illustration of the growth over-diffusion problem: numerical schemes of order 0 over-estimate the formation of large grains at low resolution. The plot has been realised with the scheme presented in \citet{Kovetz1969} for the case of a constant kernel $K = 1$ with $N=15$ logarithmically-spaced dust bins.}
\label{fig:overdiff} 
\end{figure}

Hence, 3D dust/gas simulations that include growth and fragmentation are compulsory to understand dust evolution during the early stages of planet formation (e.g. \citealt{Safronov1972,Hayashi1975,Weidenschilling1980,Ohtsuki1990,Wetherill1990,Tanaka1996,Dominik2007,Ormel2007,Birnstiel2010}). The simplest way to formalise the evolution of a local mass distribution of dust grains is by the mean of the deterministic mean-field Smoluchowski equation, which assumes binary collisions \citep{Smolu1916}. This equation does not have generic analytic solutions. \textit{Integrated non-linearities challenge numerical solvers to obtain accurate solutions} (see Fig.~\ref{fig:overdiff}). As such, this equation has been thoroughly studied since a century (e.g. \citealt{Muller1928,Schumann1940,Chandrasekhar1943,Melzak1953,McLeod1962,Golovin1963,Berry1967,Scott1968,Trubnikov1971,Hidy1972,Drake1972a,Gillespie1975b,SW1978,ST1979,Gelbard1980,Aldous1999,Friedlander2000,Ramkrishna2000,FL2004,Jacobson2005,Pruppacher2010}), and applied extensively to several fields such aerosols science, chemistry, meteorology, biology and astrophysics.

It has been shown that classical solvers require a sufficient resolution in mass to avoid artificial formation of aggregates of large masses \citep{Soong1974,Berry1974,Trautmann1999,Khain2018}. This artificial diffusion may become particularly important when the mass interval considered is large (Fig.~\ref{fig:overdiff}). Typically, for planet formation, a few hundreds of mass bins are required to compute dust growth from interstellar sizes to pebbles. Usually, this fact is of no importance given current computational capacities. However, 3D hydrodynamical simulations can hardly handle more than (a few) ten(s) of mass bins in practice. Compromises have therefore been performed either by simplifying their growth or their dynamics. However, 1-2 D hydrodynamical codes integrating the Smoluchowski equation (e.g. \citealt{Birnstiel2010}) provide different results compared to 3D hydrodynamical codes with monodisperse growth models (e.g. \citealt{Gonzalez2017}), showing the necessity of a comprehensive approach. \textit{This implies to develop a solver which solves accurately the Smoluchowski equation with a limited number of bins, tractable by 3D hydrodynamical codes.}

Reaching high accuracy with a low number of bins while conserving mass of a finite interval of mass is a characteristic property of finite volume high-order solvers, which stem therefore as a natural way to address the growth over-diffusion problem. In this study, we present a high-order solver for the Smoluchowski equation based on the Discontinous Galerkin method, following the pioneering work of \citet{Liu2019}. Important properties of the Smoluchowski equation discussed in the astrophysical context are presented in Sect.~\ref{sec:coag}. The novel Discontinous Galerkin numerical scheme is presented in Sect.~\ref{sec:dg}. The performances of the solver regarding the over-diffusion problem are studied in Sect.~\ref{sec:num}. Applicability of the algorithm to young stellar objects or in other astrophysical contexts are discussed in Sect.~\ref{sec:discussions}.

\section{Smoluchowski equation}
\label{sec:coag}

\subsection{Short summary}

The Smoluchowski equation describes mass conservation for a distribution aggregates where mass transfers are allowed. This equation exists under a discret form (monomers forming polymers) or a continuous limit form when mass quantization becomes negligible \citep{Muller1928}. The Smoluchowski equation is a non-linear integro-differential hyperbolic equation that depend on a collision function called the growth kernel (or kernel) which quantifies the collision rate between two grains. Explicit solutions exist only for the so-called constant \citep{Smolu1916,Schumann1940,Scott1968}, additive \citep{Golovin1963,Scott1968} and multiplicative kernels \citep{McLeod1962,Scott1968}, implying numerical resolution for physical problems. Among the known solutions, self-similar solutions are particularly important since they provide asymptotic behaviour of the mass distribution at large times \citep{Schumann1940,Friedlander1966,Wang1966,Menon2004,Niethammer2016a,Laurencot2018}. A generic feature of these solutions is the exponentially fast decay of the solution at large masses. Gelation, i.e. formation of aggregates of infinite mass form in a finite time for kernels sustaining explosive growth \citep{Leyvraz1981}. In astrophysics, collisions occurs essentially through ballistic impacts modulated by focusing due to long-range interactions \citep{Safronov1972,Dullemond2005}. Kernels are non-explosive and mass remains rigorously conserved during the grow process.

\subsection{Conservative form}

\begin{figure}
\includegraphics[width=\columnwidth,trim=100 200 100 100, clip]{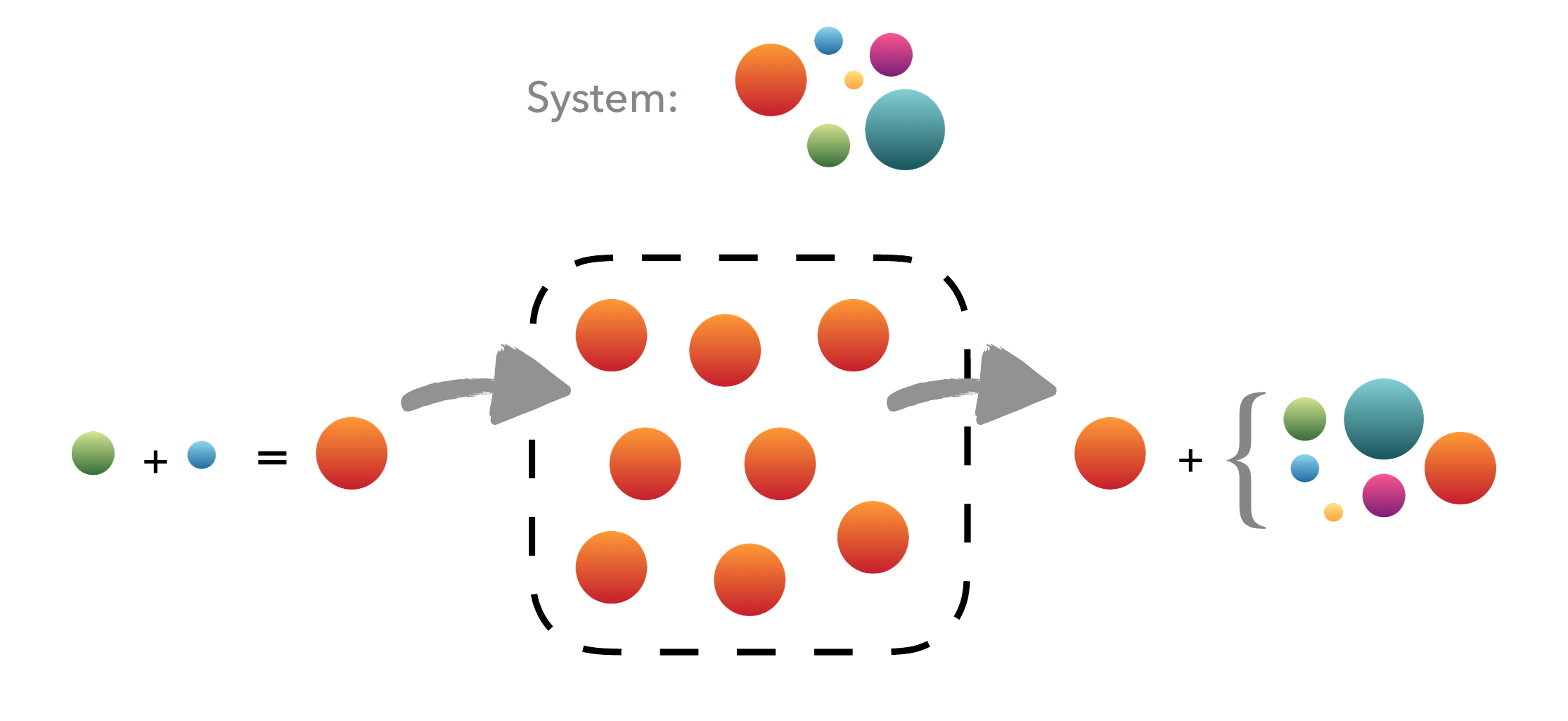}
\caption{Illustration of the Smoluchowski equation Eq.~\ref{eq:smolu_cont}. Polymers of mass $m_i$ are represented in orange. The green and blue polymers have masses lower than $m_i$. Creation (resp. growth) of polymers of mass $m_i$ increases (resp. decreases) its number density.}
\label{fig:scheme_smolu}
\end{figure}

Mass conservation for a distribution of growing grains has been originally formalised by \citet{Smolu1916}. Growth is modelled via binary collisions between spheres having known mean probabilities. The by-products of collisions are called aggregates or polymers. In \citet{Smolu1916}, aggregates are assumed to also have spherical shapes. Spatial correlations are neglected. The smallest colliding elements are referred as monomers. For physical systems involving aggregates made of large numbers of monomers, it is often convenient to assume continuous mass distributions. The population density of grains within an elementary mass range $\mathrm{d}m$ is characterised by its number density $n\!\left( m \right)$. The continuous Smoluchowski equation is given by
\begin{equation}
  \begin{aligned}
    \frac{\partial n\left(m,t\right)}{\partial t}  =  & \frac{1}{2} \int\limits_0^{m} \! K\!\left(m-m',m'\right)n \!\left(m-m',t\right)n\!\left(m',t\right) \mathrm{d}m' \\
    & -n\!\left(m,t\right) \int\limits_0^{\infty} \! K\!\left(m,m'\right)n\!\left(m',t\right) \mathrm{d}m',  
  \end{aligned}
  \label{eq:smolu_cont}
\end{equation}
where $t$ denotes time and $m$ and $m'$ the \textit{masses} of two colliding polymers. The averaged probabilities of collision are encoded inside the coagulation kernel $K\left(m,m'\right)$, which is a symmetric function of $m$ and $m'$ for binary collisions (see Sect.~\ref{sec:kernels}). Fig.~\ref{fig:scheme_smolu} shows the physical meaning of the non-linear integro-differential equation Eq.~\ref{eq:smolu_cont}. The number of grains encompassed within a given interval of masses varies since i) binary collisions of aggregates of appropriate masses can increase this population (first term of the right-hand side of Eq.~\ref{eq:smolu_cont}), but ii) those grains may themselves collide with other grains to form larger aggregates (second term of the right-hand side of Eq.~\ref{eq:smolu_cont}). This equation can be put under a convenient dimensionless form by introducing \citep{Scott1968,Drake1972a}
\begin{equation}
  \left\{ 
  \begin{aligned}
    & x \equiv m/m_0,\,y \equiv m'/m_0,\, \mathcal{K}(x,y) = K(m,m')/K_0, \\
    & \tau = (K_0 N_0) t,\, f(x,\tau) = m_0 \, n(m,t)/N_0 .
  \end{aligned}
  \right.
\end{equation}
$N_0$ is the initial total number density of particles, $m_0$ is the initial mean mass of the particles and $K_0$ is a normalising constant with dimensions $[\mathrm{length}]^3/\mathrm{time}$. We adopt the variables $x$ and $\tau$ for sake of clarity and homogeneity with the existing literature (e.g. \citealt[and references therein]{Friedlander2000,Jacobson2005}). $x$ denotes therefore masses. Eq.~\ref{eq:smolu_cont} transforms into 
 \begin{equation}
  \begin{aligned}
    \frac{\partial f(x,\tau)}{\partial \tau} = &  \frac{1}{2} \int\limits_0^x \! \mathcal{K}(y,x-y) f(y,\tau) f(x-y,\tau) \mathrm{d}y \\
    & - f(x,\tau) \int\limits_0^{\infty} \! \mathcal{K}(y,x) f(y,\tau) \mathrm{d}y.
  \end{aligned}
  \label{eq:smolu_cont_DL}
 \end{equation}
 Eq.~\ref{eq:smolu_cont_DL} is physically ill-posed, since the probability to form aggregates of mass larger than the initial mass of the system may be non-zero. Recently, \citet{Tanaka1996} have shown that Eq.~\ref{eq:smolu_cont_DL} can be equivalently written under the conservative form 
\begin{equation}
\left\{
   \begin{aligned}
   &\frac{\partial g \left( x,\tau \right) }{\partial \tau} + \frac{\partial F_{\mathrm{coag}} \left[ g \right] \left( x,\tau \right)}{\partial x} = 0 \\
   &F_{\mathrm{coag}} \left[ g \right] \left( x,\tau \right) = \int\limits_0^x \! \! \int\limits_{x-u}^{\infty} \mathcal{K} \left( u,v \right) g \left( u,\tau \right) \frac{g \left( v,\tau \right)}{v} \mathrm{d}u \mathrm{d}v ,
   \end{aligned}
\right.
\label{eq:smol_cons_DL}
\end{equation}
where $g\left(x,\tau \right) \equiv x f\left(x,\tau \right)$ is the mass density of polymers per unit mass, and $F_{\mathrm{coag}} \left[ g \right] \left( x,\tau \right)$ is the flux of mass density across the mass $x$ triggered by coagulation \citep{FL2004}. Under this conservative form, the infinite upper bound of the second integral in $F_{\mathrm{coag}}$ can simply be replaced by $x_{\rm max} - u$. This prevents the formation of aggregates of masses larger than $x_{\rm max}$ by settling the passing-through mass flux to be rigorously zero.

\subsection{Kernels}
\label{sec:kernels}

Physically, the coagulation kernel is defined according to
\begin{equation}
   K\! \left(m,m' \right) \equiv \beta\! \left(m,m', \Delta v\right) \Delta v \!\left(m,m'\right) \sigma \! \left(m,m'\right),
\end{equation}
where $\Delta v$ is the mean relative velocity between two aggregates of masses $m$ and $m'$, $\sigma$ is the mean effective cross section of collision and $\beta$ denotes the mean sticking probability of the grains. The coagulation kernel encodes the microphysics of collisions inside $\beta$, $\sigma$ and $\Delta v$, those parameters depending \textit{a priori} on the sizes of the colliding grains, or the kinetic and thermodynamical parameters of an eventual surrounding flow. A kernel of particular importance for physical problems is the Ballistic kernel (Table~\ref{table:kernels}). In this case, $\sigma$ corresponds simply to the geometric cross-section of the grains (focusing effects due to electrostatic or gravitational forces being neglected), and $\beta$ and $\Delta v$ are treated as constants (which may be a relevant approximation at least over moderate ranges of masses). Coagulation kernel can also be seen as mathematical objects useful to study the properties of the Smoluchowski equation under various conditions or to derive explicit analytic solutions. The expression of the four kernels discussed in this work is given in Table~\ref{table:kernels}.

\subsection{Analytic solutions}
\label{sec:analytic}
Explicit analytic solutions exist in the case of simple kernels and specific initial conditions. We review these solutions hereafter since they will be used in Sect.~\ref{sec:num} to benchmark the numerical algorithms.

\subsubsection{Constant kernel}
\label{sec:kconst}
For the constant kernel $\mathcal{K} \! \left( x,y \right)=1$ and the initial condition $f \left( x,0 \right) = \exp \left(-x\right)$, the solution of Eq.~\ref{eq:smolu_cont_DL} is \citep{Muller1928,Schumann1940,Melzak1957,Rajagopal1959,Scott1968,ST1979}
\begin{equation}
  \left\{
  \begin{aligned}
    &f_1(\tau) \equiv \frac{4}{(2+\tau)^2},\,f_2(\tau) \equiv \frac{\tau}{2+\tau},\\
    &f(x,\tau) = f_1(\tau) \exp\left(-\left\lbrace1-f_2(\tau)\right\rbrace x\right).\\
   \end{aligned}
   \right.
   \label{eq:sol_kconst}
\end{equation}
Physically, a constant kernel $\mathcal{K}=1$ implies that the frequency of collisions between two aggregates is independent of their size.

\begin{table}
\begin{center}
\begin{tabular}{cc}
  \hline
  Kernel & $\mathcal{K}(x,y)$ \\
  \hline
  Size-independent & $1$ \\
  Sum & $x+y$ \\
  Product & $xy$ \\
  Ballistic & $\pi\left(x^{1/3}+y^{1/3}\right)^2 \Delta v$
\end{tabular}
\caption{Functional form of the different coagulation kernels $\mathcal{K}$ considered in this study.}
\label{table:kernels}
\end{center}
\end{table}

\subsubsection{Additive kernel}
\label{sec:kadd}
The solution for the additive kernel $\mathcal{K}(x,y)=x+y$ with the initial condition $f_0(x,0)=\exp(-x)$ has been derived by \citet{Golovin1963}.  \citet{Scott1968} extended the derivation for a general initial condition. For an initial condition under the form $f(x,0) = \exp(-x)$, the solution of Eq.~\ref{eq:smolu_cont_DL} is
\begin{equation}
  \left\{
   \begin{aligned}
   & T \equiv 1-\exp(-\tau), \\
   &f(x,\tau) = \frac{\left(1-T\right)\exp\left(-x\left\lbrace1+T\right\rbrace\right)}{xT^{1/2}}I_1 \! \left(2xT^{1/2}\right),\\
   \end{aligned}
   \right.
   \label{eq:sol_kadd}
\end{equation}
where $I_1$ is the modified Bessel function of first kind. Physically, the additive kernel implies that the frequency of collisions increases according to the size of the grains. Large aggregates form faster compared to case of a constant kernel, leading to broader dust distributions at large masses. The asymptotic tail presents therefore a smoother decay compared to the case $\mathcal{K}=1$.

\subsubsection{Multiplicative kernel}
\label{sec:kmul}
Originally, \citet{McLeod1964} derived a solution for the multiplicative kernel $\mathcal{K}(x,y) = xy$ with the initial condition $f_0(x,0)=x^{-1}\exp(-x)$ only for a small interval of time. The general solution for this problem was later found by \citet{Ernst1984}
\begin{equation}
  \left\{
   \begin{aligned}
   & T \equiv \left\{
   \begin{aligned}
      & 1+\tau \quad \mathrm{if} \, \,\,\, \tau \leq 1  \\
      & 2\tau^{1/2} \quad \mathrm{otherwise}
   \end{aligned}, \right. \\
   &f(x,\tau) = \frac{\exp\left(-Tx\right) I_1 \! \left(2x \tau^{1/2} \right)}{x^2 \tau^{1/2}}.\\
   \end{aligned}
   \right.
   \label{eq:sol_kmul}
\end{equation}
The multiplicative kernel is a typical kernel to study the occurrence of gelation, since at $\tau=1$, aggregates with infinite masses form and mass conservation is mathematically no longer satisfied. Physically, the multiplicative kernel means an explosive increase of the collisional frequencies with respect to grain sizes. Massive grains form faster compared to the case of the additive kernel. In the same time, the mass density of small grains decreases quickly.

\subsection{Numerical methods}
\label{sec:numerical_methods}
No known analytic solutions exist for the Smoluchowski coagulation equation with physical kernels, implying numerical resolution. Various numerical schemes have been developed for this purpose. Two classes of algorithms have been developed. A first class of solvers consists of Monte-Carlo simulations (e.g. \citealt{Gillespie1975a,Liffman1992,Smith1998,Lee2000,Debry2003,Sheng2006,Ormel2007,Zsom2008}). Although convenient, these methods have two principal drawbacks. Firstly, a large number of particles is required to ensure appropriate accuracy of the number density distribution $f$. Secondly, the scheme is not deterministic and simulations can be reproduced only in a statistical sense, which is not satisfying when interfacing with hydrodynamics. A second class of solvers consist of  deterministic algorithms. These methods have been summarised in \cite{Kostoglou1994,Kumar1996a,Ramkrishna2000,Pruppacher2010,Khain2018}. A short but comprehensive summary is given hereafter.

\subsubsection{Method of moments}
\label{sec:moment_method}
The method of moments seems to be the first numerical method proposed to solve the Smoluchowski equation \citep{Hulburt1964}. A system of ordinary differential equations is written over the $k$th moments  $M_k \equiv \int_0^{\infty} x^k f(x,\tau) \mathrm{d}x$ of the number density function. Approximations either for the reconstruction of $f$ \citep{Hulburt1964} or for the derivation of fractional moments \citep{Estrada2008} are then required to close this system of ordinary differential equations. The Standard Moment Method (SMM) requires an analytical integration of the kernel. To avoid this difficulty, Quadrature Moment Methods (QMM), where integrals are approximated by Gaussian quadrature methods, have been developed. Solutions of moments can be used directly to derive the total number of particles $M_0$, the total mass $M_1$ or other physical quantity such as dust opacities \citep{Marchisio2003,Estrada2008}. Number densities $f$ are reconstructed using polynomials \citep{Pruppacher1980,Piskunov2002}. 

\subsubsection{Point-based methods}
\label{sec:point_based_method}
The number density function $f$ is sampled over a mass grid. The main difficulty lies in representing the continuous distribution $f$ as accurately as possible using the values of $f$ at the sampling points. Different algorithms have been developed using this approach:

\paragraph{Interpolation method} 
This method was developed by \citet{Berry1967,Berry1974}. The continuous Smoluchowski equation is written in terms of $g(x,\tau) \equiv x f(x,\tau)$, the mass density function. The mass interval is discretised using a logarithmic grid. A system of ordinary differential equations is derived with respect to the variable $g$ evaluated on the grid points. Gain and loss terms are evaluated separately, and integrals are calculated by using high-order Lagrangian interpolations. \citet{Middleton1976,Suck1979} improved this method by using Simpson's rules for the integrals and cubic splines interpolations.

\paragraph{Method of orthogonal collocation}
The method of weighted residuals \citep{Finlayson1972} is a general method for obtaining numerical solutions to differential equations. The unknown solution is tested over a set of weight functions and is adapted to give the best approximated solution to the differential equation. The Smoluchowski equation is multiplied by the weight function $\phi$ and integrated over all the mass domain to form the residual
\begin{equation}
   \begin{aligned}
      R \equiv \int_0^{\infty} & \left(\frac{\partial f(x,\tau)}{\partial \tau} \right.  - \int_0^x \mathcal{K}(x-y,y)f(x-y,\tau)f(x,\tau) \mathrm{d}y  \\
      &  \left. + \int_0^{\infty} \mathcal{K}(x,y) f(x,\tau)f(y,\tau) \mathrm{d}y \right) \phi(x) \mathrm{d}x =0.
   \end{aligned}
\end{equation}
The number density $f$ is approximated by polynomials. The collocation method corresponds to the case where $\phi(x) = \delta (x-x_0)$. The coagulation equation is evaluated at the collocation points $x_0$. This gives a set of ordinary differential equations equal to the degree of freedom of the polynomials used. Integrals are usually performed using Gaussian quadrature rules \citep{Eyre1988}.
\begin{figure}
\centering
\includegraphics[width=\columnwidth,trim=0 200 0 150, clip]{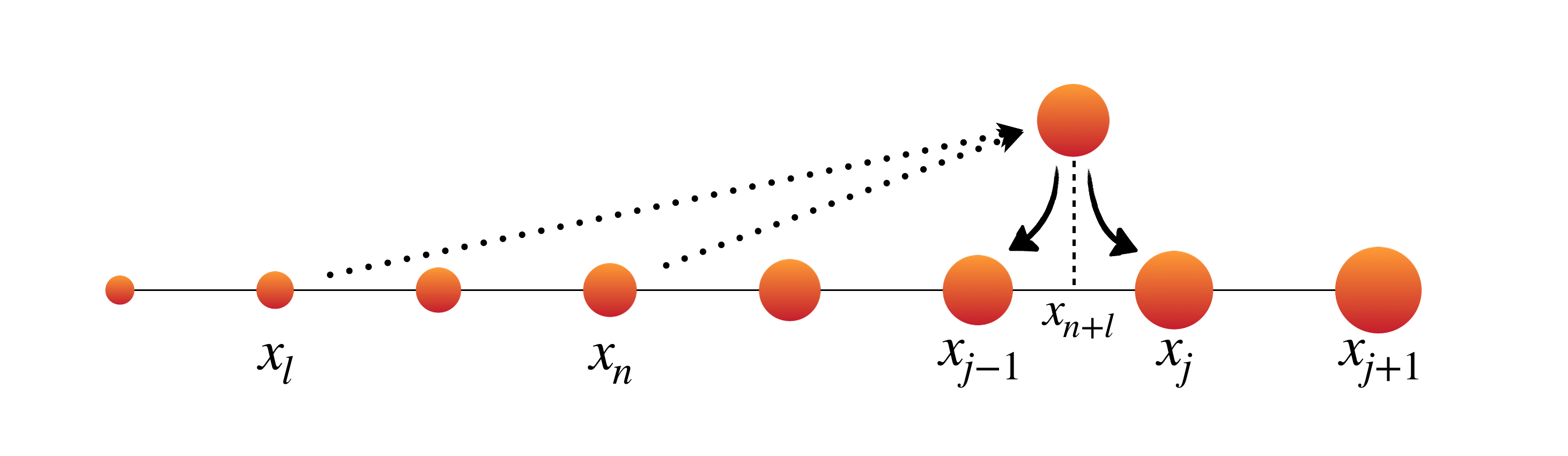}
\caption{Illustration of the pair interaction methods. A particle of mass $x_{n+l}=x_n +x_l$ forms from collision between particles of masses $x_l$ and $x_n$. The resulting mass $x_{n+l}$ is distributed onto adjacent bins, generating numerical over-diffusion towards large masses.}
\label{fig:scheme_pair_bins}
\end{figure} 

\paragraph{Pair interaction methods} Numerical integration of the Smoluchowski equation consists of summing contributions of pairwise collisions between all grid points of different masses. For non-regular mass samplings, aggregates do usually not have masses corresponding to an existing grid point. To ensure mass conservation, the mass of the aggregate is distributed over the two relevant adjacent grid points (Fig.~\ref{fig:scheme_pair_bins}). The first pair-interaction solver has been developed by \citet{Kovetz1969}. In this algorithm, a system of ordinary differential equations is obtained over the quantities $N(x_i)=\int_{a_i}^{b_i} \! f(x)\mathrm{d}x$ where $x_i$ denotes the mass of individual particles of the $i$-th point, and $a_i \equiv (x_{i+1}-x_{i})/2$ and $b_i\equiv(x_{i}-x_{i-1})/2$. In practice, logarithmic grids are used to cover wide ranges of masses. In the context of planet formation, widely used solvers follow this approach (e.g. \citealt{Brauer2008,Birnstiel2010}). The principal drawback of this method is that redistribution of mass towards large grains tend to over-predict the number of large aggregates, triggering artificial formation of large bodies (Fig.~\ref{fig:overdiff}). A large number of grid points is therefore required to avoid an artificial broadening of number density of particles $f$ \citep{Berry1974,Soong1974,Khain2018}. Moreover, a sufficient number of grid points is also needed to avoid difficulties related to collisions that form aggregates of masses larger than the largest mass point. \citet{Jacobson2005} extended the \citet{Kovetz1969} algorithm by distributing the mass between grid points and writing the scheme in a semi-implicit form. This solver ensures mass conservation to machine precision. \citet{Bott1998,Simmel2002,Wang2007} developed also binary-pairs interaction methods. Mass is advected towards adjacent grid points by a mass flux expressed with a high-order scheme. These methods do not introduce a significant numerical broadening. Other methods have been developed by \citet{Hounslow1988,Lister1995} where four binary interaction mechanisms of gain and loss of particles are considered to deal correctly the rate of change of particle and mass.

\subsubsection{Finite element methods}
\label{sec:finite_element_method}
In these methods, the continuous mass distribution is discretised over a finite number of mass elements (intervals, cells, bins).

\paragraph{Moments with finite elements} The first finite element scheme for coagulation was developed by \citet{Bleck1970} by discretising mass distributions over logarithmic bins. $f$ is approximated by its moment of order zero over each bin to obtain a system of ordinary differential equations. Over-diffusion for large grains is observed with this piecewise constant approximation. A change of variable $x \rightarrow x^{-3}$ is operated to reduce diffusivity at large masses. The method of \citet{Soong1974} follows \citet{Bleck1970}. The Smoluchowski equation is written in terms of mass density distributions $g$ and approximated by piecewise exponential functions. This allows to reduce drastically the diffusive effect at large masses. \citet{Gelbard1980,Landgrebe1990} proposed a similar method, where the Smoluchowski equation is decomposed over bins of indices $j$ in terms of $Q_j=\int_{I_j} x f(x,\tau) \mathrm{d}x$. A precise account of gain and loss of particles in terms of fluxes of $Q$ is performed. \citet{Trautmann1999} extends the work of \citet{Gelbard1980}, also finding numerical diffusion when using piecewise constant approximation, and addressing it by using piecewise exponential approximations. Another moment method that involves polynomial approximations for the first two moments $M_0$ and $M_1$ of $f$ has been proposed by \citep{Enukashvily1980,Kumar1996a,Tzivion1999}.
      
\paragraph{Discontinuous Galerkin method} The discontinuous Galerkin method is a weighted residual method where the weight $\phi(x)$ consists of orthogonal polynomials (Lagrange polynomials, Legendre polynomials, cubic splines). The numerical solution of the Smoluchowski equation is decomposed on each bin over this basis and a system of ordinary differential equations is obtained for the coefficients (e.g. \citealt{Pilinis1990,Erasmus1994,Mahoney2002}, see Sect.~\ref{sec:dg}). Generally, the integrals are performed by Gaussian quadrature rules \citep{Gelbard1978,Rigopoulos2003,Sandu2006}.     

\subsubsection{Finite element schemes in the conservative form}
\label{sec:fem_cons_form}
The conservative form Eq.~\ref{eq:smol_cons_DL} has been exploited for numerical simulations only lately. \citet{FL2004} derived a finite volume scheme of order zero where volume integrals over flux divergences are replaced by flux terms at the interfaces by the mean of the divergence theorem. This scheme conserves mass exactly and has been further extended by \citep{Filbet2008,Bourgade2008,Forestier2012}. The mass interval can be sampled uniformly or non-uniformly. Finite volume schemes of higher orders solving for the conservative form have been investigated recently \citep{Gabriel2010,Liu2019}. \citet{Gabriel2010} used WENO reconstruction \citep{Jiang2000} to approximate the coagulation flux at interfaces. \citet{Liu2019} developed a numerical scheme based on the discontinuous Galerkin method. This method provides the further advantage to choose the order of the scheme in a flexible manner. Integrals are calculated using Gaussian Quadrature rules, which implies sub-sampling of the mass intervals.

\begin{figure}
\centering
\includegraphics[width=0.9 \columnwidth]{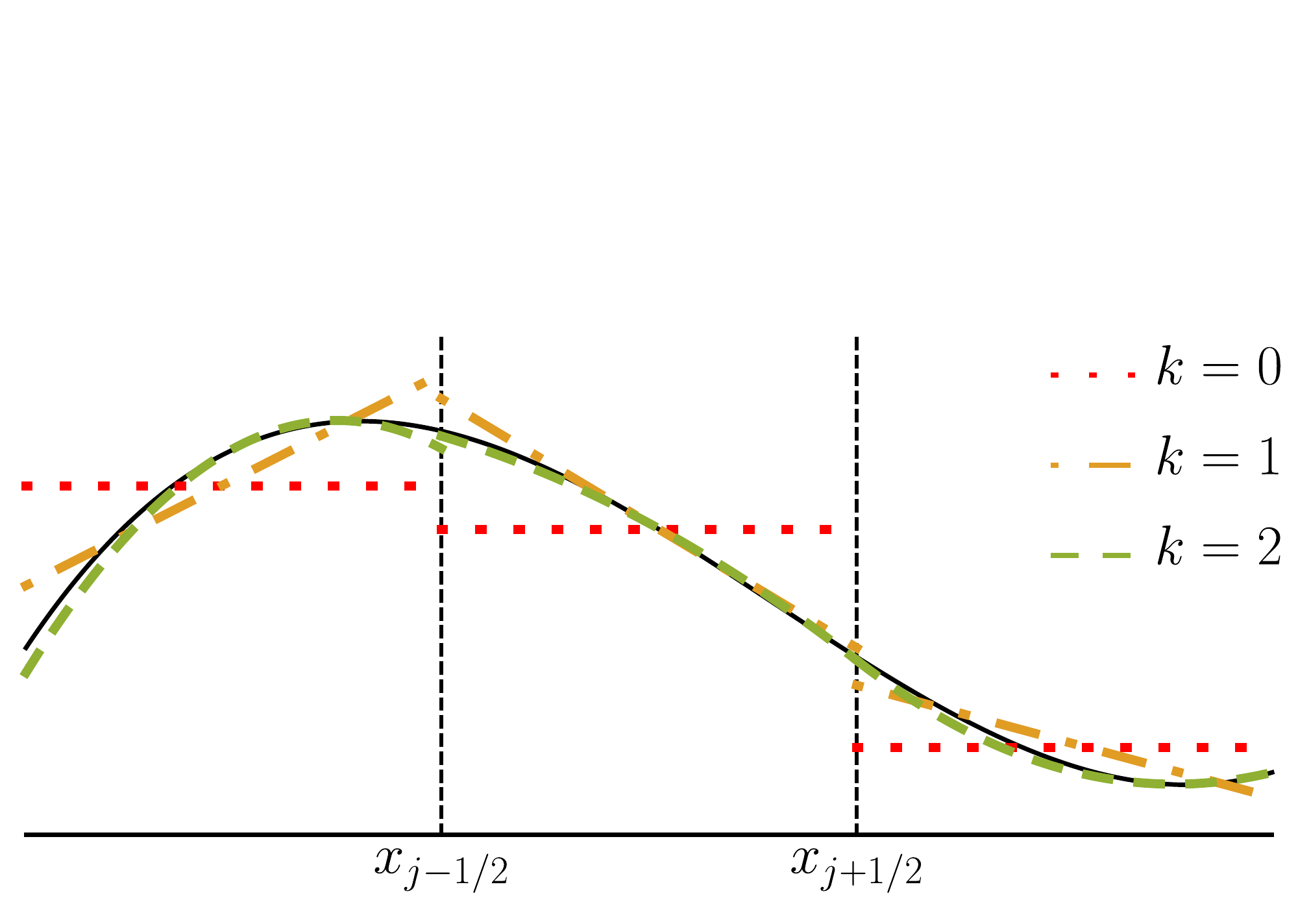}
\caption{Sketch of the discontinuous Galerkin method. In each cell, the solution is approximated by high-order polynomials $k$ to increase accuracy.}
\label{fig:scheme_DG} 
\end{figure}

\subsection{Requirements from hydrodynamical simulations}
\label{sec:hydro}

Densities must remain strictly positive and total mass conserved rigorously to ensure the stability of hydrodynamical simulations. These two properties are genuinely ensured by finite volume methods based on the conservative form Eq.~\ref{eq:smol_cons_DL}. The double-integral formulation allows to simply quench the formation of aggregates with unphysical masses, by setting for the integral bound the maximum mass allowed. These constrains may not always be satisfied with simple integral formulations.

On the other hand, observational constrains on young stellar objects are essentially provided by high-contrast spectro-polarimetry at infrared wavelengths (SPHERE/VLT, GPI, Subaru/HiCIAO) and millimetre interferometry (ALMA). These observations probe (sub)micron-to-millimetre-in-size dust distributions in discs, which corresponds to 4 orders of magnitude in size, i.e. 12 orders of magnitude in mass for compact grains. With current computational capacities, 3D dust/gas simulations of dusty discs can handle $\sim$10-20 dust species simultaneously (e.g. \texttt{PHANTOM}, \citealt{Price2018} or \texttt{RAMSES}, \citealt{Lebreuilly2020}). The global accuracy of second-order hydrodynamical solvers is of order $\sim 10^{-3}$. We aim therefore to design a versatile algorithm for coagulation of accuracy $\sim 10^{-3}$ with $\sim15$ dust bins distributed over 12 orders of magnitude in mass that allows tractable simulations. We therefore face the issue of over-diffusion associated to piecewise constant reconstructions with few mass bins, and high-order schemes appear as a natural way to overcome this difficulty. It is much preferable for hydrodynamics to handle a fix grid of sizes, to avoid interpolations when updating forces. We seek therefore for a growth algorithm that works efficiently with a fixed grid.

Additionally, we seek for an algorithm which allows for convergence analysis in 3D hydrodynamical simulations. As explained above, multiplying the number of dust bins provides prohibitive computational costs. Instead, the order of the scheme may be varied, should it be parametrised in a flexible manner. This requirement tends to favour Discontinuous Galerkin schemes with respect to WENO schemes, although they provide in theory equivalent accuracies. Compared to regular Galerkin schemes, discontinuous Galerkin solvers decompose the solution over several mass bins. This helps to better capture the exponential decay of the solution at large masses and avoid over-diffusion biases. For these reasons, we have chosen to focus on the Discontinuous Galerkin method to solve for the Smoluchowski equation in astrophysical contexts, an approach recently pioneered by \cite{Liu2019}.

Monofluid dust/gas hydrodynamical solvers provide a natural architecture to include a coagulation equation. Indeed, relative drifts between grains of different sizes are genuinely computed, eventually in the terminal velocity approximation (e.g. \citealt{Laibe2014a,Hutchison2018,Lebreuilly2019}). Monofluid formalism also ensures exact conservation of momentum, i.e. no thrust due to mass transfers propel the mixture. Sub-grid fluctuations should be prescribed by an accurate model that describes local turbulence or Brownian motion.
\section{Discontinuous Galerkin algorithm}
\label{sec:dg}

\subsection{Discontinuous Galerkin method}
\label{sec:DG_method}
The discontinuous Galerkin method is presented for the general scalar hyperbolic conservative equation
\begin{equation}
   \left\{
   \begin{aligned}
      &\frac{\partial g(x,\tau)}{\partial \tau} + \frac{\partial F[g](x,\tau)}{\partial x} =0,\\
      &(x,\tau) \in \mathbb{R}_+,
   \end{aligned}
   \right.
   \label{eq:general_cons_law_eq}
\end{equation}
where $g$ is a density of a conservative quantity and $F\!\left[g \right]$ the associated flux.\\

Let partition the domain of interest $[x_{\mathrm{min}},x_{\mathrm{max}}] \in \mathbb{R}$ in $N$ subintervals (alternatively, cells or bins), not necessarily of equal sizes. Each cell is defined by $I_j=(x_{j-1/2},x_{j+1/2}],  j \in [\![1,N]\!]$. The size of the $j$-th cell is defined as $h_j = x_{j+1/2}-x_{j-1/2}$. The cell is centred around the position $x_j=\left(x_{j+1/2}+x_{j-1/2}\right)/2$. We define $\mathcal{V}^k$ the space of polynomials of degree $k$ in each cell $I_j$
\begin{equation}
   \mathcal{V}^k = \left\{ v:v|_{I_j} \in P^k \left( I_j \right),j \in [\![1,N]\!] \right\}.
   \label{eq:basis}
\end{equation}
We denote $g_j \in \mathcal{V}^k$ the approximate solution of $g$ in the bin $I_j$. The terminology \textit{discontinuous Galerkin (DG)} comes from the fact that in $\mathcal{V}^k$, the functions are allowed to have jumps at the interfaces $x_{j+1/2}$. One obtains a weak formulation of Eq.~\ref{eq:general_cons_law_eq} 
by multiplying by a test function $\phi \in \mathcal{V}^k$, integrating over $I_j$ and finally integrating by parts \citep{Cockburn1989}
\begin{equation}
   \begin{aligned}
   \int_{I_j} \frac{\partial g_j}{\partial t} \phi \mathrm{d}x &- \int_{I_j} F[g]\left(x,t\right) \frac{\partial \phi}{\partial x} \mathrm{d}x \\
   &+ F[g]\left(x_{j+1/2},t\right) \phi(x_{j+1/2}) \\
   & - F[g]\left(x_{j-1/2},t\right) \phi(x_{j-1/2})  = 0 .
   \end{aligned}
    \label{eq:DG_eq}
\end{equation}
Eq.~\ref{eq:DG_eq} allows to fix unequivocally the degrees of freedom of the function $g_j$. The residual of Eq.~\ref{eq:general_cons_law_eq} on bin $I_j$ is defined as
\begin{equation}
   \begin{aligned}
   R_j \equiv \int_{I_j} \frac{\partial g_j}{\partial t} \phi \mathrm{d}x & - \int_{I_j} F[g]\left(x,t\right) \frac{\partial \phi}{\partial x} \mathrm{d}x \\
   & +F[g]\left(x_{j+1/2},t\right) \phi(x_{j+1/2}) \\
   & - F[g]\left(x_{j-1/2},t\right) \phi(x_{j-1/2}).
   \end{aligned}
\end{equation}

DG schemes consist of choosing a local orthogonal polynomials basis on $I_j$ to replace the test function and to approximate the solution. Residuals $R_j$ are therefore null in the sense of orthogonalisation on the basis. 
In practice, Legendre polynomials are used \citep{Cockburn1989}. We denote hereafter the $i$-th Legendre polynomial by $\phi_i\left(\xi\right)$, where $\xi \in [-1,1]$. Polynomial functions $\phi_i\left(\xi\right)$ are orthogonal in $L^2\left([-1,1]\right)$ with respect to the inner product with weight unity. Fig.~\ref{fig:scheme_DG} shows a sketch of the DG method. In each cell, the function $g$ is approximated by Legendre polynomials. The accuracy of the approximation increases with respect to the order of the polynomials. The approximation of $g$ in cell $I_j$ writes
\begin{equation}
  \begin{aligned}
    & \forall x \in I_j,\; g(x) \approx g_j\left(x,t\right)=\sum_{i=0}^k g_j^i\left(t\right) \phi_i(\xi_j\left(x\right)), \\
    & g_j\left(x,t\right)= \bm{g}^T_j(t) \cdot \bm{\phi}(\xi_j (x)), \; \mathrm{with} \;
    \bm{g}_j =
    \begin{bmatrix} 
    g_j^0 \\ \vdots \\  g_j^k
    \end{bmatrix} \mathrm{and} \;
    \bm{\phi} =
    \begin{bmatrix} 
    \phi_0 \\ \vdots \\ \phi_k
    \end{bmatrix},
  \end{aligned}
  \label{eq:proj_basis}
\end{equation}
where $g_j^i$ is the component of $g_j$ on the Legendre polynomials basis. The function $\xi_j\left(x\right) \equiv \frac{2}{h_j}\left(x-x_j\right)$ is used to map the interval $I_j$ onto the interval $[-1,1]$. Normalising the Legendre basis gives
\begin{equation}
  \int\limits_{-1}^1 \bm{\phi}(\xi) \bm{\phi}^T (\xi) \mathrm{d}\xi = d_i \delta_{ik} \; \mathrm{with} \;d_i \equiv \frac{2}{2i+1},
  \label{eq:normalisation_leg}
\end{equation} 
where $d_i$ is the coefficient of normalisation. By combining Eqs.~\ref{eq:DG_eq}, \ref{eq:proj_basis} and \ref{eq:normalisation_leg} one obtains
\begin{equation}
  \begin{aligned}
    &\frac{\mathrm{d} \bm{g}_j\left(t \right)}{\mathrm{d} t}  = \bm{L}[g] \; \mathrm{with} \\
    &  
    \begin{aligned}
      \bm{L}[g] \equiv & \frac{2}{h_j} 
        \begin{bmatrix}
          1/d_0 & & \\ & \ddots & \\ & & 1/d_i
        \end{bmatrix} \\ 
      & 
      \begin{aligned}
        \Bigg( \Bigg.  \int_{I_j}  & F\left[ g \right]\left( x,t \right) \partial_x \bm{\phi} \left(\xi_j\left(x \right)\right) \mathrm{d}x \\ 
        & 
        \begin{aligned}
          -  \bigg[ \bigg. &  F\left[ g \right]\left(x_{j+1/2},t\right) \bm{\phi} \left(\xi_j \left( x_{j+1/2} \right) \right)   \\
          & \Bigg. \bigg. -  F\left[ g \right]\left(x_{j-1/2},t\right) \bm{\phi} \left(\xi_j \left( x_{j-1/2} \right) \right)  \bigg]  \Bigg) ,
        \end{aligned} 
      \end{aligned} 
    \end{aligned}
  \end{aligned}
  \label{eq:DG_ode}
\end{equation}
where $\bm{L}$ is the operator that results from applying the DG procedure to Eq.~\ref{eq:general_cons_law_eq} with a Legendre polynomials basis. With the procedure described above, the original system of partial differential equations (PDE) Eq.~\ref{eq:DG_eq} is transformed into a system of ordinary differential equations (ODE) Eq.~\ref{eq:DG_ode} onto the coefficients $g_j^i(t)$. The initial condition $g_j\left(x,0\right)$ is generated by the piecewise $L^2$ projection of an initial mass density distribution $g_0(x)$ on each bin, i.e.
\begin{equation}
  \begin{aligned}
    & \forall j \in [\![1,N]\!], \\
    & \int_{I_j} \left( g_j\left(x,0\right) - g_0\left(x\right) \right) \bm{\phi}^T (\xi_j(x)) \mathrm{d}x = \bm{0}. 
  \end{aligned}
\end{equation} 
Orthogonality of Legendre polynomials ensures
\begin{equation}
  \begin{aligned}
  \int_{I_j} g_j \bm{\phi}^T \mathrm{d}x & = \frac{h_j}{2} \int_{-1}^1 \bm{\phi}(\xi) \bm{\phi}^T (\xi) \mathrm{d}\xi  \bm{g}_j (t) \\
  & = \frac{h_j}{2} \mathrm{diag}[d_0,...,d_k]  \bm{g}_j (t).
  \end{aligned}
\end{equation}
Then, the components of $\bm{g}_j$ are given by
\begin{equation}
  \begin{aligned}
  &\forall j \in [\![1,N]\!], \forall i \in [\![0,k]\!],\\
  &g_j^i(0) = \frac{2}{h_j d_i} \int\limits_{-1}^1 g_0 \left(\frac{h_j}{2} \xi_j + x_j \right) \phi_i (\xi_j) \mathrm{d} \xi_j.
  \end{aligned}
  \label{eq:g_components_initial}
\end{equation}
Hence, the DG method consists in solving the following Cauchy problem
\begin{equation}
   \left\{
   \begin{aligned}
      &\forall j \in [\![1,N]\!], \forall i \in [\![0,k]\!], \\
      &\frac{\mathrm{d} \bm{g}_j\left(t \right)}{\mathrm{d} t}  = \bm{L}[g],\\
      &g_j^i(0) = \frac{2}{h_j d_i} \int\limits_{-1}^1 g_0 \left(\frac{h_j}{2} \xi_j + x_j \right) \phi_i (\xi_j) \mathrm{d} \xi_j,
   \end{aligned}  
   \right.
\end{equation}
where $\bm{L}$ is detailed in Eq.~\ref{eq:DG_ode}.

\subsection{Evaluation of the flux}
\label{sec:fluxes}

\subsubsection{Regularised flux}

The continuous Smoluchowski coagulation Eq.~ \ref{eq:smolu_cont_DL} is defined over an unbounded interval of masses $x \in \mathbb{R}_+$. Before applying the DG procedure, Eq.~\ref{eq:smolu_cont_DL} is restrained to a physical mass interval. Moreover, growth from a gaseous reservoir is excluded, meaning that $x>0$. The mass interval is therefore reduced to the interval $[x_{\mathrm{min}} > 0,x_{\mathrm{max}} < + \infty]$ \citep{FL2004,Liu2019}. The coagulation flux can be truncated according to two procedures \citep{FL2004}. On the one hand 
\begin{equation}
  \begin{aligned}
    & F_{\mathrm{coag}}^{\mathrm{c}} \left[ g \right] \left( x,\tau \right) = \\
    & \qquad \int\limits_{x_{\mathrm{min}}}^x \! \int\limits_{x-u+x_{\mathrm{min}}}^{x_{\mathrm{max}}-u+x_{\mathrm{min}}} \mathcal{K} \left( u,v \right) g \left( u,\tau \right) \frac{g \left( v,\tau \right)}{v} \mathrm{d}v \mathrm{d}u,
   \end{aligned}
\end{equation}
where $F_{\mathrm{coag}}^{\mathrm{c}}$ is the conservative flux, meaning that no particle of mass larger than $x_{\mathrm{max}}$ is allowed to form. On the other hand
\begin{equation}
   F_{\mathrm{coag}}^{\mathrm{nc}} \left[ g \right] \left( x,\tau \right) = \int\limits_{x_{\mathrm{min}}}^x \! \int\limits_{x-u+x_{\mathrm{min}}}^{x_{\mathrm{max}}} \mathcal{K} \left( u,v \right) g \left( u,\tau \right) \frac{g \left( v,\tau \right)}{v} \mathrm{d}v \mathrm{d}u,
\end{equation}
where $F_{\mathrm{coag}}^{\mathrm{nc}}$ is the non-conservative flux which allows formation of particles of mass $x>x_{\mathrm{max}}$. $F_{\mathrm{coag}}^{\mathrm{c}}$ is useful in realistic simulations of growth, whereas $F_{\mathrm{coag}}^{\mathrm{nc}}$ should be used to compare numerical solution to analytic solutions of Eq.~\ref{eq:smolu_cont}. \\

\subsubsection{Method for evaluating the flux}
A crucial difference between this scheme and usual DG solvers is that the coagulation flux $F_{\mathrm{coag}}^{\mathrm{nc}}$ is non local. The evaluation of the numerical flux $F_{\mathrm{coag}}^{\mathrm{nc}}[g]$ at the interface $x_{j+1/2}$ depends on the evaluation of $g_j$ in all cells. Mathematically, $F_{\mathrm{coag}}^{\mathrm{nc}}$ is a double integral of a product of polynomials. Then the flux is a continuous function of mass $x$. At the interface $x_{j+1/2}$, the numerical flux reduces to $F_{\mathrm{coag}}^{\mathrm{nc}}\left[ g \right] = F_{\mathrm{coag}}^{\mathrm{nc}}\left[ g \right] \left(x_{j+1/2},t\right)$. In usual DG solvers, the numerical flux is a discontinuous function and must be reconstructed at the interfaces (e.g \citealt{Cockburn1989,Zhang2010}).

The principal difficulty lies in carefully evaluating the flux at interfaces. This relies on handling the numerical integration of the polynomials $g_j$ in every relevant cell. \cite{Liu2019} uses a Gaussian quadrature method with a Legendre polynomials basis to approximate the flux. The lower bound of the inner integral $x-u$ does usually not correspond to a grid point. To accurately perform the Gauss quadrature, some grid elements must be sub-divided, increasing drastically the cost of the numerical procedure, especially for high-order polynomials. To avoid prohibitive computational costs due to cell oversampling, we take advantage of the polynomial approximation by calculating integrals analytically. This requires integrable kernels, which is the case for the four kernels presented in this study. This approach maintains a reasonable computational cost by not multiplying the number of sampling points. This also avoid to add errors due to the numerical integration and to approximate kernels by piecewise constant functions.

\subsubsection{Mathematical procedure}

To integrate analytically the numerical flux, let define the function $\tilde{g}$ that approximates the function $g$ over the entire mass interval
\begin{equation}
  \begin{aligned}
    &\forall x \in [x_{\mathrm{min}},x_{\mathrm{max}}],\\
    &\tilde{g}\left(x,\tau \right) \equiv \\
    & \sum_{l=1}^N \sum_{i=0}^k g_l^i\left(\tau\right) \phi_i(\xi_l(x)) [ \theta(x-x_{l-1/2}) - \theta(x-x_{l+1/2})].
  \end{aligned}
\end{equation}
We assume that the kernel function is explicitly integrable and can be written as $\mathcal{K}(u,v) = \mathcal{K}_1(u) \mathcal{K}_2(v)$, which is effectively the case for the three simple kernels and the ballistic kernel (see Sect.~\ref{sec:kernels}). For instance, the additive kernel writes $\mathcal{K}_{\mathrm{kadd}}(u,v) = u+v = \mathcal{K}_1^1(u) \mathcal{K}_2^1(v) + \mathcal{K}_1^2(u) \mathcal{K}_2^2(v)$, where $\mathcal{K}_1^1(u)=u$, $\mathcal{K}_2^1(v)=1$, $\mathcal{K}_1^2(u)=1$ and $\mathcal{K}_2^2(v)=v$. The numerical flux is split in two terms. The numerical flux writes
\begin{equation}
  \begin{aligned}
  & F_{\mathrm{coag}}^{\mathrm{nc}}[\tilde{g}](x,t) = \sum_{l'=1}^{N} \sum_{i'=0}^k \sum_{l=1}^{N} \sum_{i=0}^k g_{l'}^{i'}(t) g_l^i(t)\\
  &
    \begin{aligned}
      \int\limits_{x_{\mathrm{min}}}^x  \int\limits_{x-u+x_{\mathrm{min}}}^{x_{\mathrm{max}}} & \frac{\mathcal{K}(u,v)}{v} \\
      & \phi_{i'}(\xi_{l'}(u)) [\theta(u-x_{l'-1/2})-\theta(u-x_{l'+1/2})] \\
      & \phi_{i}(\xi_l(v)) [\theta(v-x_{l-1/2})-\theta(v-x_{l+1/2})] \mathrm{d}v \mathrm{d}u.
    \end{aligned}
  \end{aligned}
  \label{eq:numerical_flux}
\end{equation}
In the DG Eq.~\ref{eq:DG_eq}, the numerical flux is evaluated on grid points $x_{j+1/2}$ and $x_{j-1/2}$  with $j \in [\![1,N]\!]$. $k$ is the order of the Legendre polynomials to approximate the solution. Therefore, $F_{\mathrm{coag}}^{\mathrm{nc}}$ depends on $j$ and $k$. The flux is sampled over a $2$D array $(N,k+1)$ in order to use vectorial operations to reduce the computational time. The numerical flux is
\begin{equation}
   \left\{
   \begin{aligned}
      & F_{\mathrm{coag}}^{\mathrm{nc}}[\tilde{g}](x,t) = \\
      & \sum_{l'=1}^{N} \sum_{i'=0}^k \sum_{l=1}^{N} \sum_{i=0}^k g_{l'}^{i'}(t) g_l^i(t) T(x,x_{\mathrm{min}},x_{\mathrm{max}},i',i,l',l),\\
      & T(x,x_{\mathrm{min}},x_{\mathrm{max}},i',i,l',l) = \\
      &
      \begin{aligned}
         &\int\limits_{x_{\mathrm{min}}}^x  \mathcal{K}_1(u)\phi_{i'}(\xi_{l'}(u)) [\theta(u-x_{l'-1/2})-\theta(u-x_{l'+1/2})] \\
         & \int\limits_{x-u+x_{\mathrm{min}}}^{x_{\mathrm{max}}} \frac{\mathcal{K}_2(v)}{v} \phi_{i}(\xi_l(v)) \\
         & \qquad \qquad [\theta(v-x_{l-1/2})-\theta(v-x_{l+1/2})] \mathrm{d}v \mathrm{d}u.
      \end{aligned}
   \end{aligned}
   \right.
   \label{eq:numerical_flux_split}
\end{equation}
\textit{A priori}, the boundaries for the intervals of integration can be arbitrarily large. We therefore rescale these intervals to avoid any numerical issues related to large numbers when calculating the terms $T$ in the variables $\xi_{l'}$ and $\xi_l$. To avoid critical typos, the term $T$ is derived with \textsc{Mathematica} by starting with the inner integral on $\xi_l$ and then the integral on $\xi_{l'}$. Further details about the derivation of the algorithm are given in supplementary material on GitHub (see Data Availability Sect.~\ref{data_github}) for reproducibility. These integrals do not commute. The high-order solver is written in \texttt{Fortran}. Reducing the number of integrals is key to avoid numerical issues with differences of large numbers. For this purpose, the expression of $T$ is split in several terms provided on GitHub (see Data Availability Sect.~\ref{data_github}). For robustness, all these integrals are calculated with \textsc{Mathematica}. The \textsc{Mathematica} function \texttt{FortranForm} is used to translate integral expressions to \texttt{Fortran}. For large expressions, it is necessary to split them with the function \texttt{MonomialList}. The scheme to evaluate $T(x,x_{\mathrm{min}},x_{\mathrm{max}},i',i,l',l)$ in \texttt{Fortran} is given on GitHub (see Data Availability Sect.~\ref{data_github}).

A $4\mathrm{D}$ array with element $T(x,x_{\mathrm{min}},x_{\mathrm{max}},i',i,l',l)$ and a $4\mathrm{D}$ array with element $g_{l'}^{i'}(t) g_l^i(t)$ are computed. The element $(j,k)$ of the $2$D array corresponding to the flux is obtained by multiplying these two $4\mathrm{D}$ arrays and summing over of all elements. $F_{\mathrm{coag}}^{\mathrm{nc}}[\tilde{g}]$ is then evaluated in $x_{j-1/2}$ and $x_{j+1/2}$ for all $j$.

\subsection{Evaluation of the integral of the flux}
\label{sec:intflux}
Let denote $\mathcal{F}_{\mathrm{coag}}^{\mathrm{nc}}$ the term of Eq.~\ref{eq:DG_eq} corresponding to the integral of the numerical flux. $\mathcal{F}_{\mathrm{coag}}^{\mathrm{nc}}$ writes
\begin{equation}
   \left\{
  \begin{aligned}
  & \mathcal{F}_{\mathrm{coag}}^{\mathrm{nc}} [\tilde{g},j,k](t) = \\
  & \sum_{l'=1}^{N} \sum_{i'=0}^k \sum_{l=1}^{N} \sum_{i=0}^k \, g_{l'}^{i'}(t) \, g_l^i(t) \mathcal{T}\left( x_{\mathrm{min}},x_{\mathrm{max}},j,k,i',i,l',l\right)\\
  & 
    \begin{aligned}
      &\mathcal{T}\left( x_{\mathrm{min}},x_{\mathrm{max}},j,k,i',i,l',l\right) \equiv \\
      &\int\limits_{I_j} \int\limits_{x_{\mathrm{min}}}^x  \int\limits_{x-u+x_{\mathrm{min}}}^{x_{\mathrm{max}}} \frac{\mathcal{K}(u,v)}{v} \, \partial_x \phi_k (\xi_j (x)) \\
      & \qquad  \phi_{i'}(\xi_{l'}(u)) [\theta(u-x_{l'-1/2})-\theta(u-x_{l'+1/2})] \\
      & \qquad \phi_{i}(\xi_l(v)) [\theta(v-x_{l-1/2})-\theta(v-x_{l+1/2})] \mathrm{d}v\, \mathrm{d}u\, \mathrm{d}x.
    \end{aligned}
  \end{aligned}
  \right.
  \label{eq:numerical_intflux_split}
\end{equation}
$\mathcal{F}_{\mathrm{coag}}^{\mathrm{nc}}[\tilde{g}]$ is evaluated similarly to the flux. A triple integral is derived with \textsc{Mathematica} with the changes of variables 
\begin{equation}
   \xi_l \equiv \frac{2}{h_l}\left(v-x_l\right),\, \xi_{l'} \equiv \frac{2}{h_{l'}}\left( u-x_{l'}\right),\, \xi_j \equiv \frac{2}{h_j}\left(x-x_j\right).
\end{equation}
To derive tractable equations for the integrals involving Heaviside distributions, we start to compute integrals over the variable $\xi_l$, then calculating the integral over $\xi_{l'}$ and finally, over $x$. The details of the calculations and the scheme in \texttt{Fortran} to evaluate $\mathcal{T}\left( x_{\mathrm{min}},x_{\mathrm{max}},j,k,i',i,l',l\right)$ are given in supplementary material on GitHub (see Data Availability Sect.~\ref{data_github}) for completeness. $\mathcal{F}_{\mathrm{coag}}^{\mathrm{nc}}$ is computed as a product of 4D arrays similarly to $F_{\mathrm{coag}}^{\mathrm{nc}}$. Accuracy on $T$ and $\mathcal{T}$ depends only the quality of the polynomial approximation of $g$ by $\tilde{g}$, since the integrals corresponding to $F_{\mathrm{coag}}^{\mathrm{nc}}[\tilde{g}]$ and $\mathcal{F}_{\mathrm{coag}}^{\mathrm{nc}}[\tilde{g}]$ are calculated analytically. 

\subsection{Slope limiter}
\label{sec:slope_limiter}
For most of astrophysical kernels, the solution of the Smoluchowski coagulation equation has been mathematically shown to decays with an exponential tail in at large masses \citep{Schumann1940,Menon2004}. This part is challenging to approximate with polynomials, and numerical estimates $g_j$ of $g$ in the bin $I_j$ can lead to negative values, which is not acceptable physically.\\

To preserve the positivity of solution, the requirement $g_j(x,t) \geq 0$ for $x \in I_j$ needs to be enforced. The idea is to use a scaling limiter which controls the maximum/minimum of the reconstructed polynomials \citep{Liu1996,Zhang2010,Liu2019}. This is achieved by a reconstruction step based on cell averaging. Let us consider the polynomials $g_j(x)$ of order $k$ that approximates $g(x)$ on $I_j$. Let denote $m$ and $M$ two positive reals $M_j\equiv \underset{x \in I_j}{\mathrm{max}}\, g_j(x)$, $m_j \equiv \underset{x \in I_j}{\mathrm{min}}\, g_j(x)$ and define the scaled polynomials
\begin{equation}
  \begin{aligned}
  & p_j \left(x\right) \equiv \gamma_j \left(g_j(x) - \overline{g}_j \right) + \overline{g}_j, \\
  &\gamma_j = \mathrm{min} \left\{ \left|\frac{M-\overline{g}_j}{M_j-\overline{g}_j}\right|,\left| \frac{m-\overline{g}_j}{m_j-\overline{g}_j}\right|,1 \right\} .
  \end{aligned}
\end{equation} 
where $\overline{g}_j$ refers to the cell average of $g$ in $I_j$
\begin{equation}
  \overline{g}_j \equiv \frac{1}{h_j} \int_{I_j} g_j(x,t) \mathrm{d}x.
\end{equation}
For all $j$, we assume $\overline{g}_j \in [m,M]$. $p_j(x)$ is a polynomial of order $k$ such as $\overline{p}_j=\overline{g}_j$. \citet{Liu1996} proved that $\forall x \in I_j,\; p_j(x) \in [m,M]$. This scaling limiter allows to build a maximum-principle-satisfying DG scheme, in the sense that the numerical solution never goes out of the range $[m,M]$. The main difficulty is to ensure the property $\overline{g}_j \in [m,M]$ during the evolution without loosing high accuracy. \\

In the DG scheme given by Eq.~\ref{eq:DG_ode}, polynomials $g_j(x)$ are replaced by the scaled polynomials $p_j(x)$ such as 
\begin{equation}
  \begin{aligned}
    p_j \left(x\right) & = \gamma_j \left(g_j(x) - \overline{g}_j \right) + \overline{g}_j \\
    & =  \sum_{i=0}^k \gamma_j g_j^i(t) \phi_{1,i}(\xi_j(x)) + \sum_{i=0}^k g_j^i(t) \phi_{2,i} (\xi_j(x)) 
  \end{aligned}
  \label{eq:scaling_polynomials}
\end{equation} 
with
\begin{equation}
  \begin{aligned}
    & \phi_{1,i}(\xi_j(x)) \equiv \left(\phi_i(\xi_j(x))- \frac{1}{2} \int_{I_j} \phi_i(\xi_j(x)) \mathrm{d}x \right), \\
    & \phi_{2,i}(\xi_j(x)) \equiv \frac{1}{2} \int_{I_j} \phi_i(\xi_j(x)) \mathrm{d}x.
  \end{aligned} 
\end{equation} 
Replacing $g_j$ by $p_j$ in Eq.~\ref{eq:numerical_flux_split} gives four terms for the function $T$: $T_{11}[\phi_{1,i'}\phi_{1,i}]$, $T_{12}[\phi_{1,i'}\phi_{2,i}]$, $T_{21}[\phi_{2,i'}\phi_{1,i}]$ and $T_{22}[\phi_{2,i'}\phi_{2,i}]$. For each term, a corresponding coefficient $g_{l',i'}(t)g_{l,i}(t)$ is associated, namely $\gamma_{l'}g_{l',i'}(t)g_{l,i}(t)$, $\gamma_{l'} g_{l',i'}(t)g_{l,i}(t)$, $\gamma_l g_{l',i'}(t)g_{l,i}(t)$ and $g_{l',i'}(t)g_{l,i}(t)$ (no $\gamma$ in the last term). $F_{\mathrm{coag}}^{\mathrm{nc}}$ is evaluated by summing over those four terms. The same procedure is applied for $\mathcal{F}_{\mathrm{coag}}^{\mathrm{nc}}$. Therefore, the positivity of $\tilde{g}$ is ensured in each cell.

\subsection{High-order time stepping}
\label{sec:time_solver}

\subsubsection{CFL condition}
\label{sec:timestepping}
Forward Euler discretisation of Eq.~\ref{eq:DG_eq} gives
\begin{equation}
   \begin{aligned}
      & \overline{g}_j^{n+1} = \\
      & \overline{g}_j^n - \frac{\Delta t}{\Delta x_j} \left[ F_{\mathrm{coag}}^{\mathrm{nc}}\left[ g_j \right] \left(x_{j+1/2},t\right)-F_{\mathrm{coag}}^{\mathrm{nc}}\left[ g_j \right] \left(x_{j-1/2},t\right)\right],
   \end{aligned}
   \label{eq:DG_discrete_mean}
\end{equation}
for the $n$-th time step. The Courant-Friedrichs-Lewy condition (CFL) of the scheme is chosen to guarantee the positivity of the cell average $\overline{g}_j^{n+1}>0$ \citep{FL2004}, i.e. 
\begin{equation}
  \Delta t < \frac{\Delta x_j \overline{g}_j^n }{| F_{\mathrm{coag}}^{\mathrm{nc}}\left[ g_j \right] \left(x_{j+1/2},t\right)-F_{\mathrm{coag}}^{\mathrm{nc}}\left[ g_j \right] \left(x_{j-1/2},t\right)|}.
  \label{eq:CFL_euler}
\end{equation}
This CFL condition associated with the slope limiter (see Sect.~\ref{sec:slope_limiter}) ensures the positivity of the global scheme. The CFL condition is initially dominated by small grains and softens as grains grow.

\subsubsection{Strong Stability Preserving Runge-Kutta method}
\label{sec:SSP}

In Eq.~\ref{eq:smol_cons_DL}, the spatial derivative $\partial_x F_{\mathrm{coag}}[g]$ is approximated by the nonlinearly stable operator $-L[g]$ given in Eq.~\ref{eq:DG_ode}. For hyperbolic conservation laws, nonlinear stability is characterised by the total variation diminishing (TVD) semi-norm 
\begin{equation}
  TV\left(g \right) \equiv \sum_j |\overline{g}_{j+1} - \overline{g}_j|.
\end{equation}
The spatial discretisation $-L[g]$ has the property that the total variation of the numerical solution does not increase for a forward Euler integration
\begin{equation}
  g^{n+1} = g^n + \Delta t L[g], \; \Delta t \leq \Delta t_{\mathrm{FE}},
\end{equation}
when $\Delta t_{\mathrm{FE}}$ the CFL condition determined in Eq.~\ref{eq:CFL_euler}, i.e. $TV\left( g^{n+1} \right) \leq TV\left( g^n \right)$. TVD property can be generalised to high-order time discretisation with a Strong Stability Preserving (SSP) scheme \citep{Shu1988,Gottlieb2001,Zhang2010,Liu2019}. The method is SSP if the following condition holds 

\begin{equation}
   TV\left( g^{n+1} \right) \leq TV\left( g^n \right),
\end{equation} 
and the timestep satisfies 
\begin{equation}
  \Delta t_{\mathrm{SSP}} \leq c \Delta t_{\mathrm{FE}},
\end{equation}
where $c$ is a positive coefficient. Stability arguments are based on convex decomposition of high-order methods in term of the first-order Euler elements. This ensures that SSP preserves high-order accuracy in time for any convex functional (e.g. $TV$). In practice, errors are dominated by mass discretisation. We use a SSP Runge-Kutta (SSPRK) third-order method \citep{Gottlieb2009,Zhang2010,Liu2019} which writes, with $c=1$,
\begin{equation}
  \begin{aligned}
    \bm{g}_j^{(1)} &= \bm{g}_j^n + \Delta t_{\mathrm{SSP}}\bm{L}[g_j^n], \\
    \bm{g}_j^{(2)} &= \frac{3}{4} \bm{g}_j^n +\frac{1}{4} \left( \bm{g}_j^{(1)} + \Delta t_{\mathrm{SSP}} \bm{L}[g_j^{(1)}]\right), \\
    \bm{g}_j^{n+1} &= \frac{1}{3} \bm{g}_j^n + \frac{2}{3} \left( \bm{g}_j^{(2)} + \Delta t_{\mathrm{SSP}} \bm{L}[g_j^{(2)}] \right).
  \end{aligned}
  \label{eq:SSPRK3}
\end{equation}
This SSPRK third-order method ensures that $\overline{g}_j \in [m,M]$ for $(m,M) \in \mathbb{R}_+$ at all times. Hence, under a suitable CFL condition, SSP high-order time discretisation preserves the property $\overline{g}_j \in [m,M]$ of the DG scheme and the linear scaling presented in Sect.~\ref{sec:slope_limiter} satisfies a maximum principle.

\subsection{Algorithm flowchart} Associating SSPRK with a DG scheme provides overall an high-order scheme that maintains overall a uniform high-order accuracy of the solution \citep{Zhang2010,Liu2019}. We use the SSPRK of third order given by Eq.~\ref{eq:SSPRK3}. Splitting the algorithm into the following steps ensures positivity:
\begin{enumerate}
  \item Initialisation: From the initial data $g_0(x)$, 
  \begin{enumerate}
    \item generate $\forall j \in [\![1,N]\!],\; g_j(x,0) \in \mathcal{V}^k$ by piecewise $L^2$ projection and get the components on Legendre basis Eq.~\ref{eq:g_components_initial}, \\
    \item define $[m,M]$ for which $\overline{g}_j(x,0) \in [m,M]$,
    \item replace $g_j$ by $p_j$,
  \end{enumerate}
  
  \item Evolution: Use the scheme Eq.~\ref{eq:SSPRK3} to compute $\forall j \in [\![1,N]\!], \forall i \in [\![1,k]\!], \;(g_j^i)^{n+1}$,\\

  \item Reconstruction: Use Eq.~\ref{eq:scaling_polynomials} to reconstruct $p_j(x,t)$.
\end{enumerate}

\section{Numerical tests}{}
\label{sec:num}

The high-order solver presented in Sect.~\ref{sec:dg} is benchmarked against the analytical solutions presented in Sect.~\ref{sec:analytic}, similarly to \citet{Liu2019}. Accuracy tests are performed with a small number of bins, consistently with hydrodynamical requirements.

\subsection{Error measurements}
\label{sec:errors}
Numerical simulations are carried out to i) investigate the \textit{experimental order of convergence} (EOC, \citealt{Kumar2014,Liu2019} ) , and ii) determine the efficiency of the algorithm. Relative errors are measured using a continuous norm and a discrete norm. The $L^1$ norm is a natural choice for equations of conservation. The continuous $L^1$ norm can be approximated by using a high order Gaussian quadrature rule
\begin{equation}
   \begin{aligned}
      \left\|f\right\|_1 & \equiv \int_{x_{\mathrm{min}}}^{x_{\mathrm{max}}} |f(x)| \mathrm{d}x \\
      & = \sum_{j=1}^{N} \int_{I_j} |f(x)| \mathrm{d}x \approx \sum_{j=1}^N \frac{h_j}{2} \sum_{\alpha=1}^R \omega_{\alpha} |f(x_j^{\alpha})|,
   \end{aligned}
   \label{eq:norm_L1}
\end{equation}
where $N$ is the number of bins, $h_j$ is the size of bin $I_j$, $\omega_{\alpha}$ are the weights and $x_j^{\alpha}$ are the corresponding Gauss points in cell $I_j$. We use $R=16$ for sufficient accuracy. The numerical error $e_{\mathrm{c},N}$ is measured with the continuous $L^1$ norm as
\begin{equation}
   e_{\mathrm{c},N}(\tau) \equiv \sum_{j=1}^N \frac{h_j}{2} \sum_{\alpha=1}^R \omega_{\alpha} |g_j(x_j^{\alpha},\tau) - g(x_j^{\alpha},\tau)| ,
   \label{eq:errL1_continuous}
\end{equation}
where $g$ and $g_j$ are the analytic and the numerical solutions of the Smoluchowski equation. Eq.~\ref{eq:errL1_continuous} is computed with \textsc{Mathematica} using $16$ digits for sufficient precision. The discrete $L^1$ norm is defined by evaluating $g_j$ and $g$ at the geometric mean $\hat{x}_j\equiv \sqrt{x_{j-1/2}x_{j+1/2}}$ of the bin $I_j$. The numerical error measured with this discrete $L^1$ norm is
\begin{equation}
   e_{\mathrm{d},N}(\tau) \equiv \sum_{j=1}^N h_j |g_j(\hat{x}_j,\tau) - g(\hat{x}_j,\tau)|.
   \label{eq:errL1_discrete}
\end{equation}

We follow \citet{Liu2019} to calculate the \textit{experimental order of convergence} (EOC) 
\begin{equation}
   \mathrm{EOC} \equiv \frac{\ln\left( \frac{e_N(\tau)}{e_{2N}(\tau)}\right)}{\ln(2)},
   \label{eq:EOC}
\end{equation}
where $e_N$ is the error evaluated for $N$ cells and $e_{2N}$ for $2N$ cells. For the calculation of the EOC, the numerical errors are calculated at time $\tau =0.01$ for the order of convergence of the DG scheme not to be altered by time stepping errors. The moments of the numerical solution are defined according to
\begin{equation}
   \begin{aligned}
      M_{p,N}\left(\tau\right) & = \int\limits_{x_{\mathrm{min}}}^{x_{\mathrm{max}}} x^{p-1} \tilde{g}(x,\tau) \mathrm{d}x \\
      & = \sum_{j=1}^N \int_{I_j} x^{p-1} g_j(x,\tau) \mathrm{d}x \\
      & = \sum_{j=1}^N \sum_{i=0}^k g_j^i(\tau) \int_{I_j} x^{p-1} \phi_i\left(\xi_j (x) \right) \mathrm{d}x. 
   \end{aligned}
\end{equation}
The total mass of the system writes 
\begin{equation}
      M_{1,N}(\tau) = \sum_{j=1}^N \sum_{i=0}^k g_j^i(\tau) \frac{h_j}{2} \underbrace{\int\limits_{-1}^1 \phi_i\left(\xi_j \right) \mathrm{d}\xi_j}_{= \delta_{00}=2} =  \sum_{j=1}^N h_j g_j^0(\tau) .
\end{equation}
Absolute errors on moments are given by
\begin{equation}
   e_{M_{p,N}}(\tau) \equiv \frac{|M_{p,N}(\tau) - M_p(\tau)|}{M_p(\tau)},
\end{equation}
where $M_p(\tau)$ is the moment of order $p$ at time $\tau$ for the exact solution. In usual convergence tests, errors are normalised with respect to the number of degrees of freedom of the algorithm. This is not the case here, since we compare absolute gains for the purpose interfacing it with an hydrodynamical solver.

\subsection{Practical implementation of the tests}
\label{sec:benchmark_coag}
Numerical tests are performed by comparing numerical solutions the constant, additive and multiplicative kernels to the solutions given in Eqs.~\ref{eq:sol_kconst}, \ref{eq:sol_kadd} and \ref{eq:sol_kmul}. Solutions are integrated over the intervals $x \in [10^{-3},10^6]$ for the constant and the additive kernels, and $x \in [10^{-3},10^3]$ for the multiplicative kernel. Tests are performed with \textsc{Fortran}, errors are calculated with \textsc{Mathematica} at machine precision. Quadruple precision is required for the additive kernel with $k = 2$, and for all kernels with $k = 3$. The results are shown for Legendre polynomials of order $k = 0, 1, 2, 3$. Above order $3$, numerical errors due to arithmetics of large numbers are not negligible anymore. A safety coefficient of $1/2$ is applied on the CFL condition, i.e. the coagulation time-step used in practice is $\mathrm{d}\tau_{\mathrm{coag}} = 1/2 \, \mathrm{d}\tau_{\mathrm{CFL}}$. Initial conditions are set to satisfy the analytic solution at initial time $\tau=0$. The analytical and numerical solutions are compared when particles of large masses are formed at final times $\tau$ that depend on the kernels. Simulations are performed by dividing $\tau$ into constant dumps of value $\mathrm{d}\tau $ (300 for the constant and the additive kernels, 10000 for the multiplicative kernel). Each dump is subdivided in several coagulation steps satisfying the CFL condition. The analytical derivation of the coagulation flux allows the algorithm to be efficient, i.e. to reach desired accuracy with a low computational time. To quantify efficiency, the computational time is compared to the one obtained with the scheme of \citet{Liu2019} with a number of Gauss points $Q=k+1$ on a simulation in double precision with $N=20$ bins, $k=1$ for the additive kernel and $k=2$ for the constant and multiplicative kernels. The Liu scheme is implemented by following the description of \citet{Liu2019} step-by-step, without additional optimisations. Simulations are performed in sequential on an Intel Core i$7$ $2.8\mathrm{GHz}$. We use the \texttt{gfortran v9.2.0} compiler. Such a comparison is delicate to perform and interpret, since it is implementation-dependant. Should the number of Gauss points in the Liu algorithm be increased to better approximate the integral terms calculated here analytically, this may result in an increase of computational time by several orders of magnitudes, giving the false impression that the Liu algorithm is not performant. Hence the choice $Q=k+1$. Qualitatively, our scheme is more effective by a factor of several unities for same precision and without requiring sub-binning, except for the additive kernel for which the Liu scheme exhibits serendipitous super-convergence \citep{Liu2019}.

\subsection{Constant kernel}
\label{sec:kconst_tests}

\subsubsection{Positivity and mass conservation}
\label{sec:kconst_positivity}
\begin{figure}
\centering
\includegraphics[width=\columnwidth]{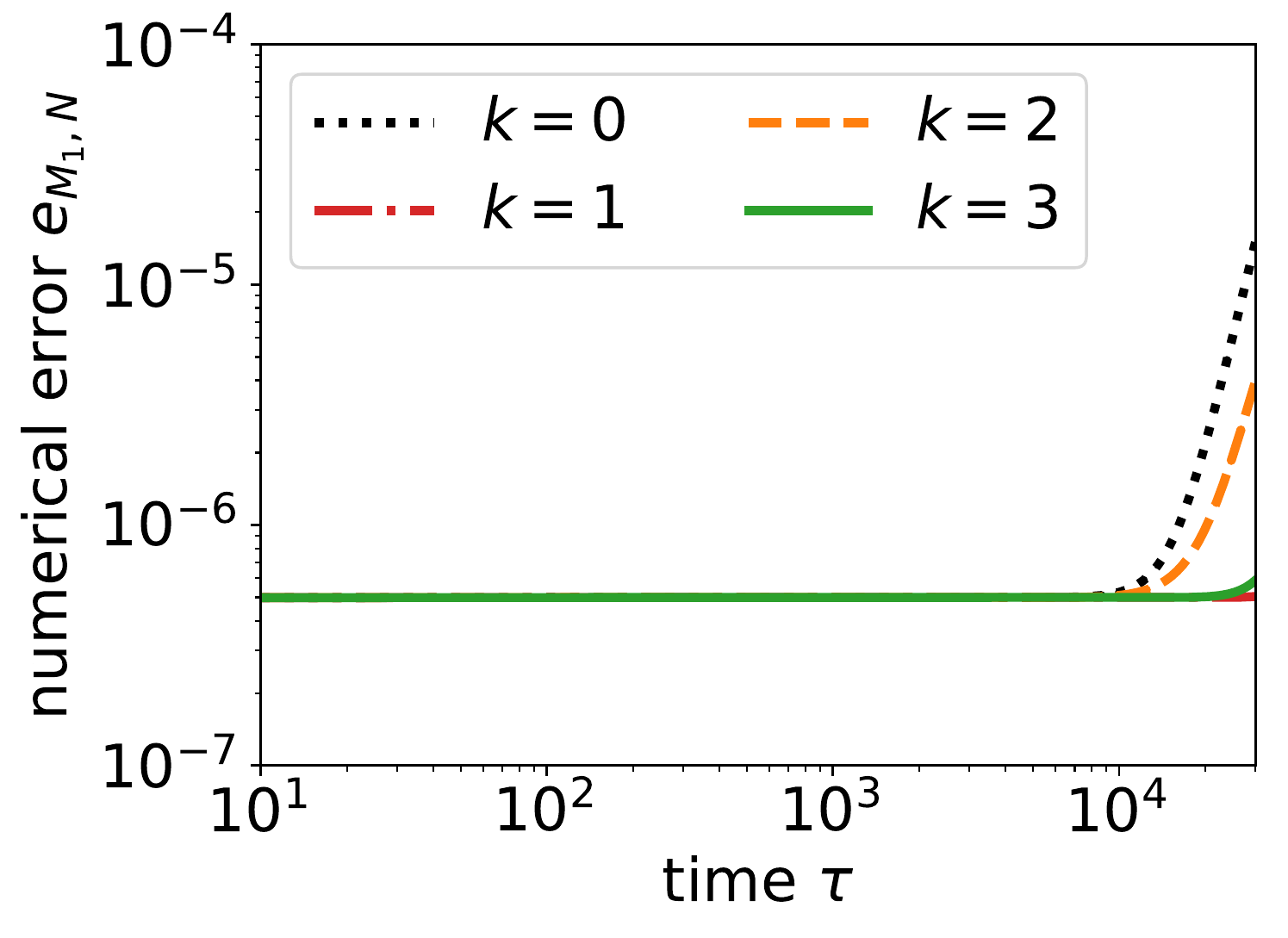}
\caption{Test case, constant kernel: evolution of the numerical absolute error $e_{M_1,N}$ on the moment $M_{1,N}$ for $N=20$ bins. The divergence at long times is explained by accumulation of errors due to numerical diffusion for even orders $k = 0$ and $k = 2$. Total mass is conserved at machine precision until $\tau=10^4$.}
\label{fig:mass_cons_kconst}
\end{figure}

\begin{figure*}
\centering
\includegraphics[width=0.8\textwidth]{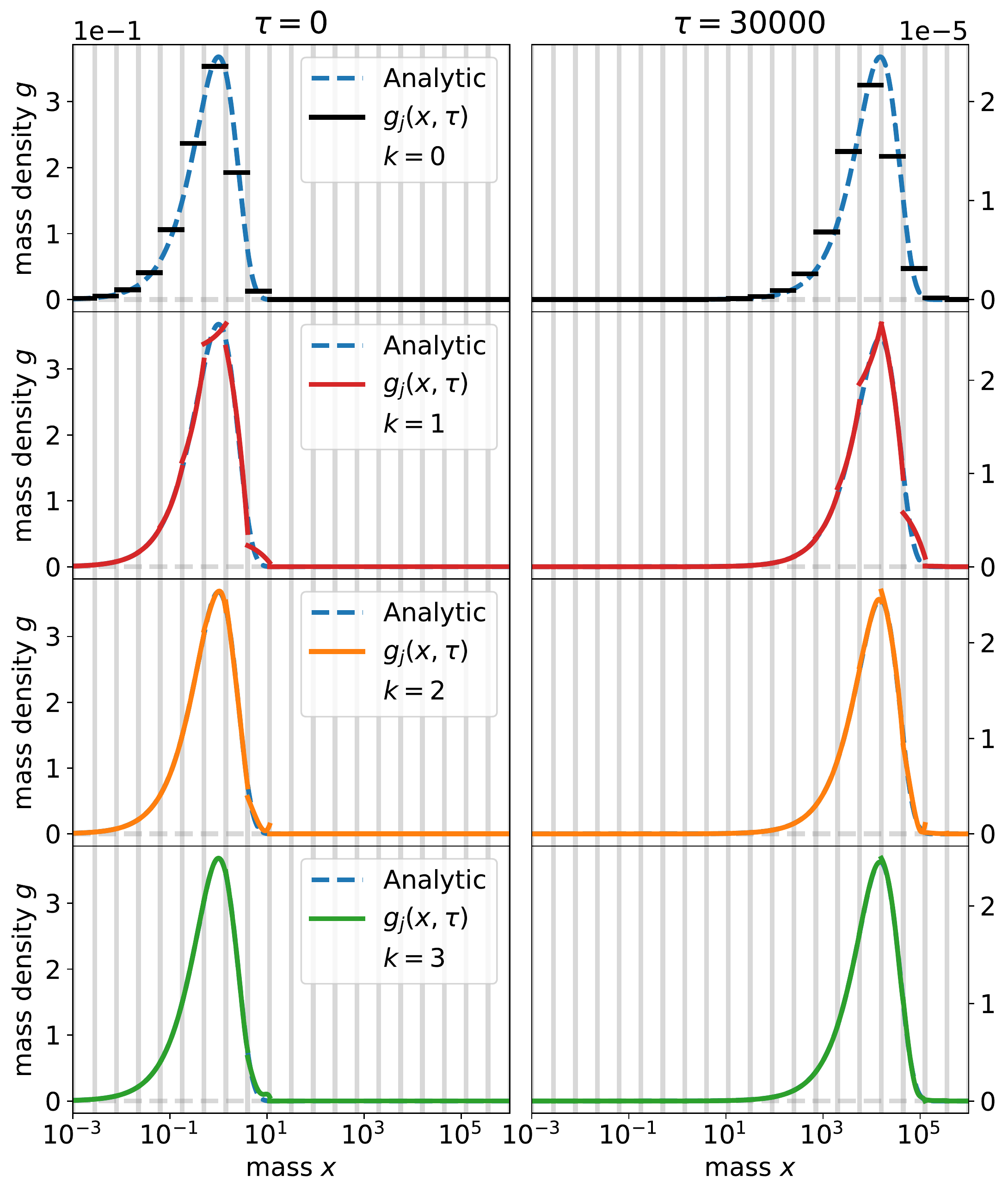}
\caption{Test case, constant kernel: the numerical solution $g_j(x,\tau)$ is plotted for $N=20$ bins and $k = 0,1,2,3$ from $\tau=0$ to $\tau=30000$, and compared to the analytic solution $g(x,\tau)$. Vertical grey lines delimit the bins. The accuracy improves for larger values of $k$. Order $3$ approximates the bump where the major part of the mass is concentrated with accuracy of order $\sim 0.1 \%$.}
\label{fig:kconst_linlog}
\end{figure*}

Fig.~\ref{fig:kconst_linlog} shows the numerical solutions obtained for $N=20$ bins, varying the order of the polynomials $k$. The analytical and numerical solutions are compared at time $\tau=30000$. As expected, the solution remains positive, as a result from combining the slope limiter (see Sect.~\ref{sec:slope_limiter}) and the SSP Runge-Kutta time stepping (see Sect.~\ref{sec:SSP}). The piecewise linear solution ($k=1$) appears curved due to the logarithmic scale of the $x$-axis.  Fig.~\ref{fig:mass_cons_kconst} shows the numerical absolute error $e_{M_1,N}$ on the moment $M_{1,N}$ for $N=20$ bins from $\tau=0$ to $\tau=30000$. The total mass remains conserved to machine precision until $\tau=10^4$.\\
\subsubsection{Accuracy of the numerical solution}
\label{sec:kconst_accuracy}
\begin{figure}
\centering
\includegraphics[width=\columnwidth]{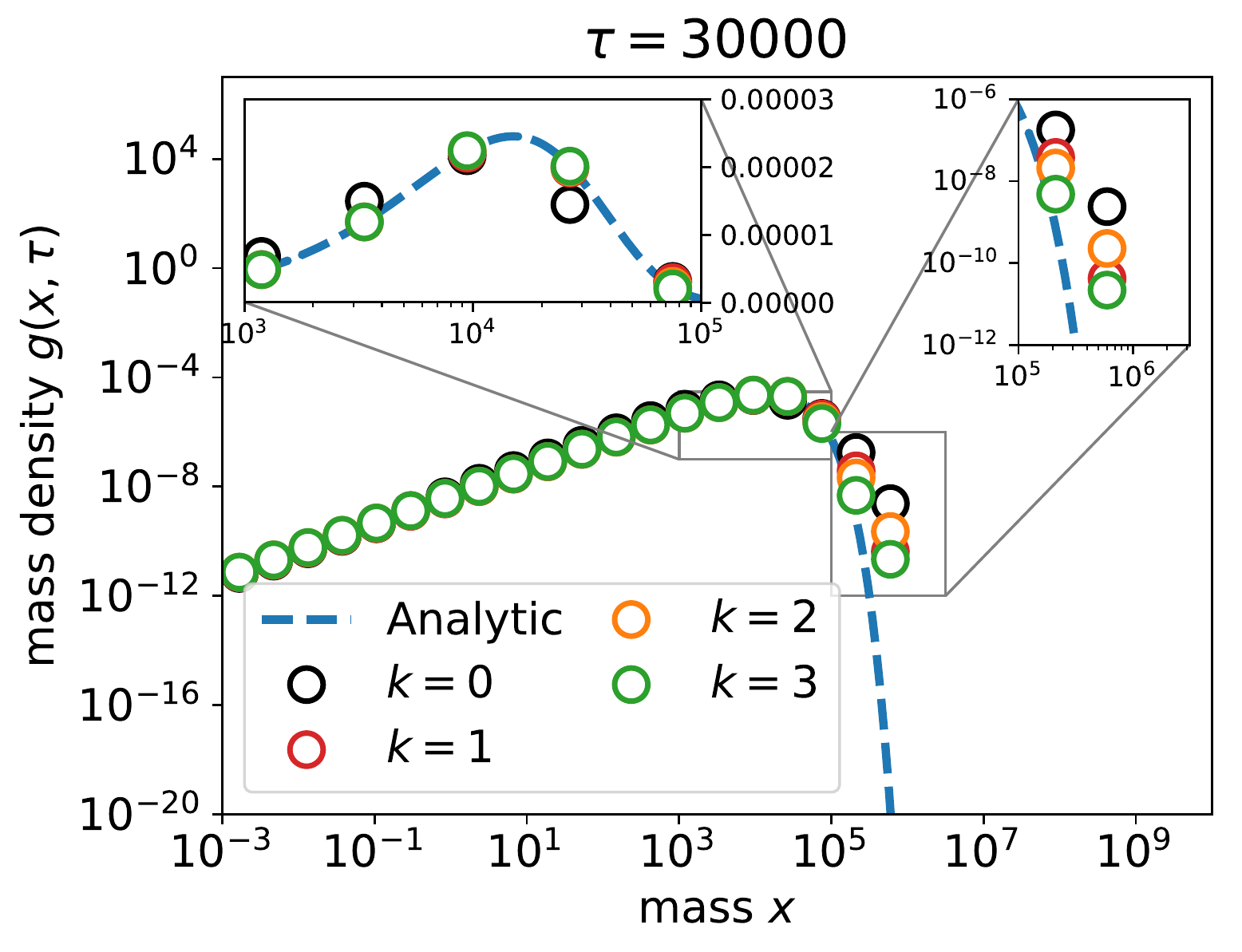}
\caption{Test case, constant kernel: numerical solution $g_j(x,\tau$) evaluated with the geometric mean $\hat{x}_j$ over each bin $I_j$. At the location of the maximum, orders $k = 1,2,3$ achieve an absolute error of $\sim 0.1 - 1\%$, to be compared with $30\%$ obtained with $k = 0$. Accuracy in the exponential tail is improved by a factor $100$ with $k = 3 $ compared to $k = 0$.}  
\label{fig:kconst_loglog_xmeanlog}
\end{figure}

As expected, the accuracy of the numerical solution improves with the order of the scheme. Fig.~\ref{fig:kconst_loglog_xmeanlog} shows the numerical solution obtained at $\tau=30000$ (note the \textit{16 orders of magnitude in mass} on the $y$ axis in log). The major part of the total mass of the system is located around the maximum of the curve. Fig.~\ref{fig:kconst_loglog_xmeanlog} shows that around this maximum, schemes of order $k = 1,2,3$ provide errors of order $\sim 0.1 - 1\%$ when $k = 0$ generates errors of order $\sim 30\%$. Fig.~\ref{fig:kconst_loglog_xmeanlog} also shows that numerical diffusion is drastically reduced in the exponential tail as the order of the scheme increases, since a gain of a factor $\sim 100$ is obtained with order $3$ compared to order $0$.

\subsubsection{Convergence analysis}
\label{sec:kconst_convergence}
\begin{figure}
\centering
\includegraphics[width=\columnwidth]{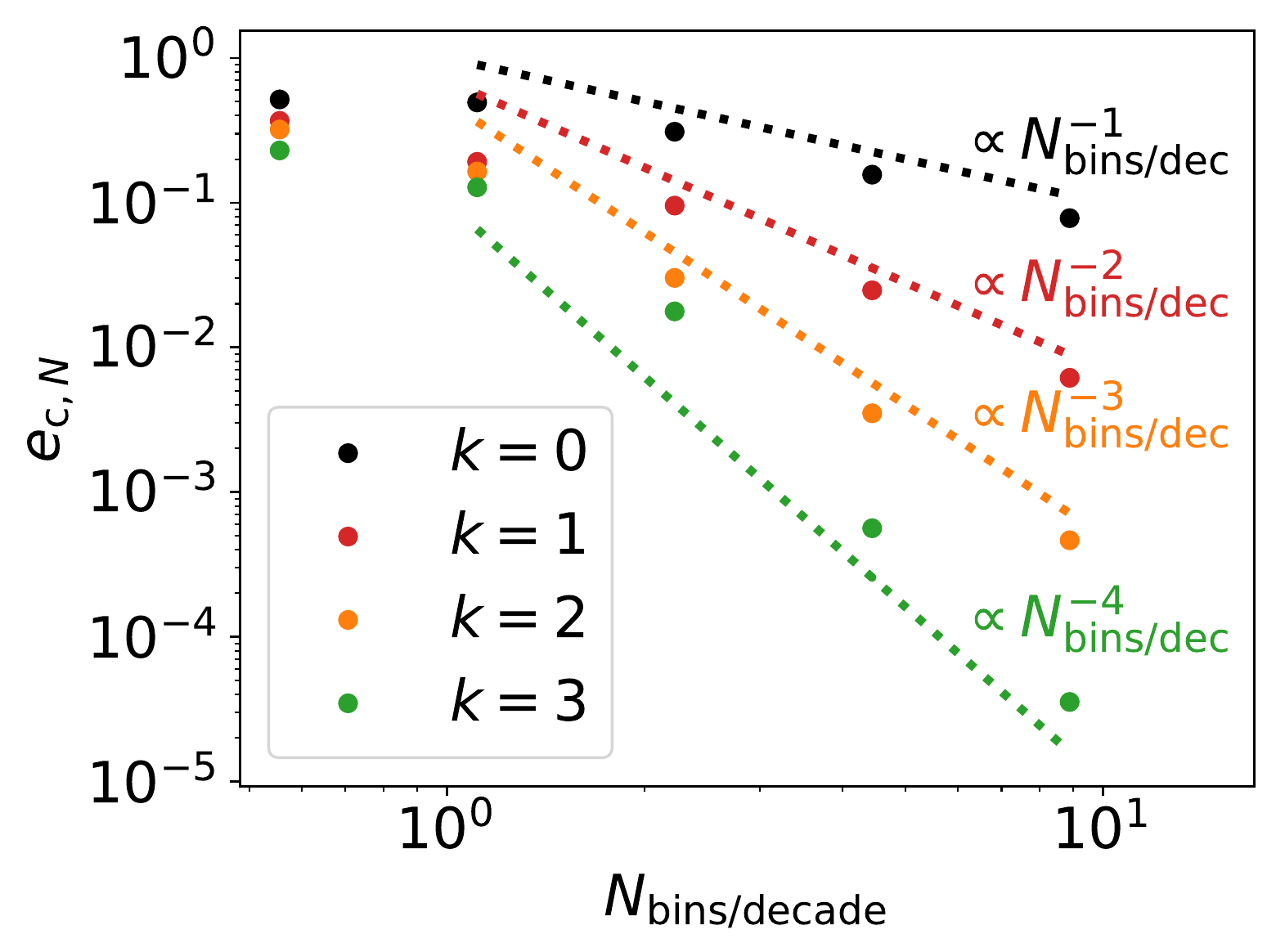}
\includegraphics[width=\columnwidth]{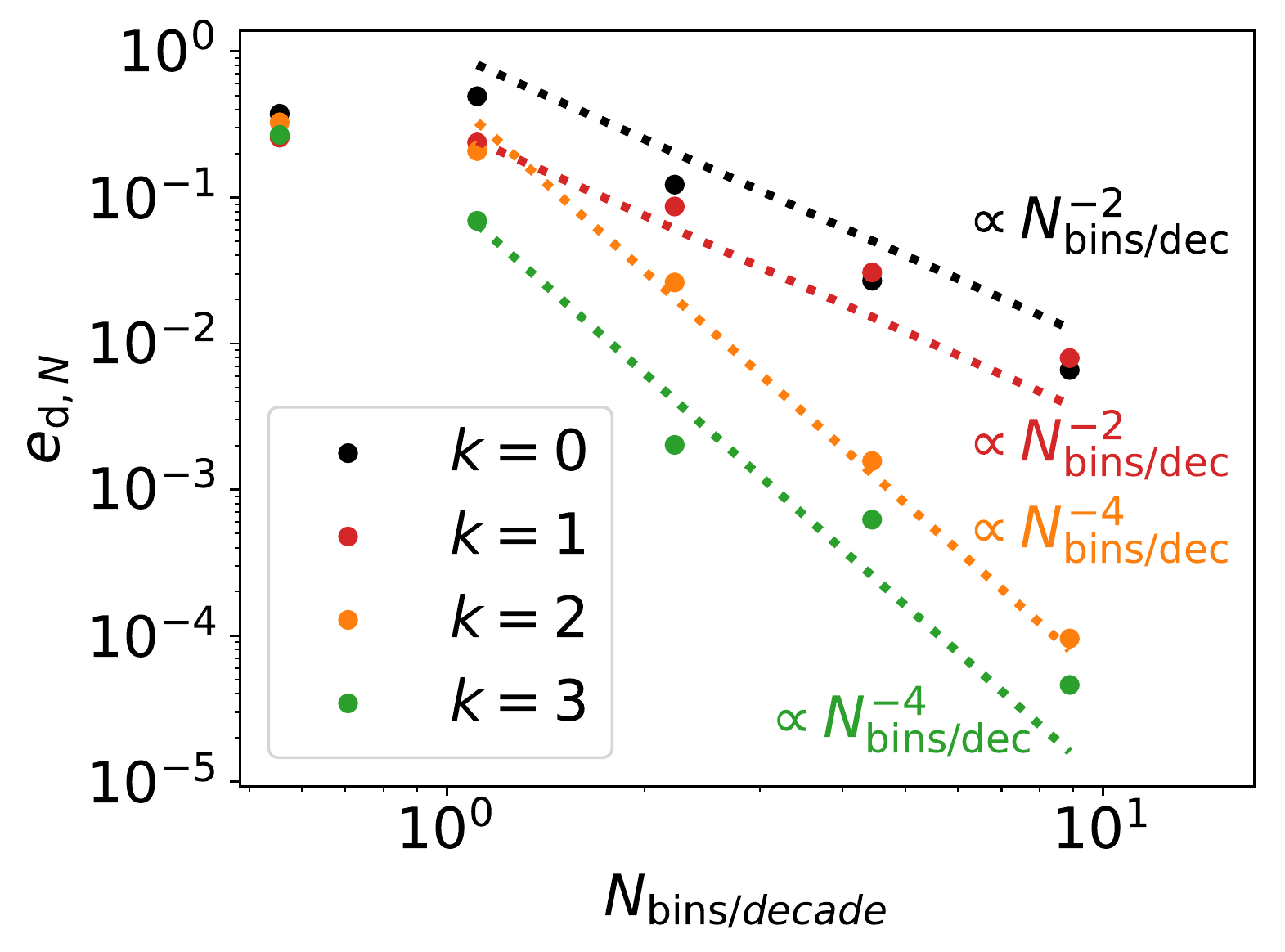}
\caption{Test case, constant kernel: the continuous $L^1$ error $e_{c,N}$ and the discrete $L^1$ error $e_{d,N}$ are plotted as functions of the number of bins per decade. With $e_{c,N}$, the experimental order of convergence is EOC = $k+1$. With $e_{d,N}$, EOC = $k+1$ for polynomials of odd orders and EOC = $k+2$ for polynomials of even orders. The DG scheme achieves on $e_{d,N}$ an accuracy of $0.1\%$ with more than $10$ bins/decade for $k=0,1$, with $\sim 9$ bins/decade for $k = 2$ and with $\sim 5$ bins/decade for $k = 3$. An accuracy of $1\%$ is achieved with $\sim 9$ bins/decade for $k = 0,1$, with $\sim 5$ bins/decade for $k = 2$ and $\sim 2$ bins/decade for $k = 3$.}
\label{fig:kconst_errL1_convergence}
\end{figure}

Numerical errors introduced in Sect.~\ref{sec:errors} are shown on Fig.~\ref{fig:kconst_errL1_convergence} at $\tau=0.01$. $e_{\mathrm{c},N}$ and $e_{\mathrm{d},N}$ are plotted as a functions of the number of bins per decade $N_{\mathrm{bin}/\mathrm{dec}}$, to infer the EOC independently from the global mass interval. With the continuous $L^1$ norm, the EOC is of order $k+1$ on a geometric grid, similarly to \citet{Liu2019}. With the discrete $L^1$ norm, the EOC is of order $k+2$ for odd polynomials, and $k+1$ for even polynomials. We recover second order of convergence (EOC=$2$) for the finite volume scheme with $k = 0$ that was predicted by \citet{FL2004}. Fig.~\ref{fig:kconst_errL1_convergence} shows that the expected accuracy of order $\sim 0.1\%$ on $e_{d,N}$ is achieved with more than $10$ bins/decade for orders $0$ and $1$, with $\sim 9$ bins/decade for order $2$ and with $\sim 5$ bins/decade for order $3$. Accuracy of order $\sim 1\%$ is achieved with $\sim 9$ bins/decade for orders $0$ and $1$, with $\sim 5$ bins/decade for order $2$, and with $\sim 2$ bins/decade for order $3$. 

\subsubsection{Stability in time}
\label{sec:kconst_stability_time}
\begin{figure}
\centering
\includegraphics[width=\columnwidth]{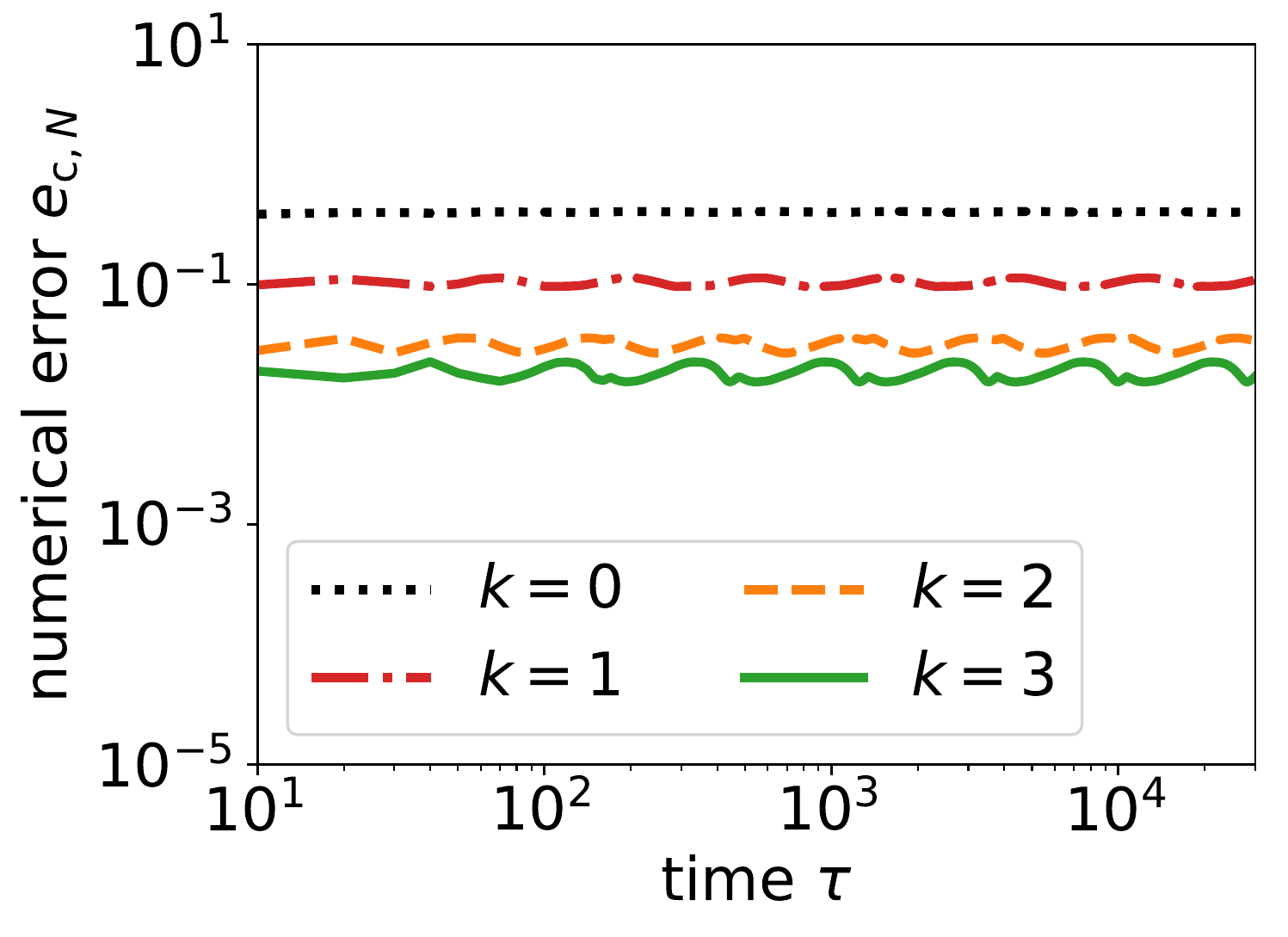}
\includegraphics[width=\columnwidth]{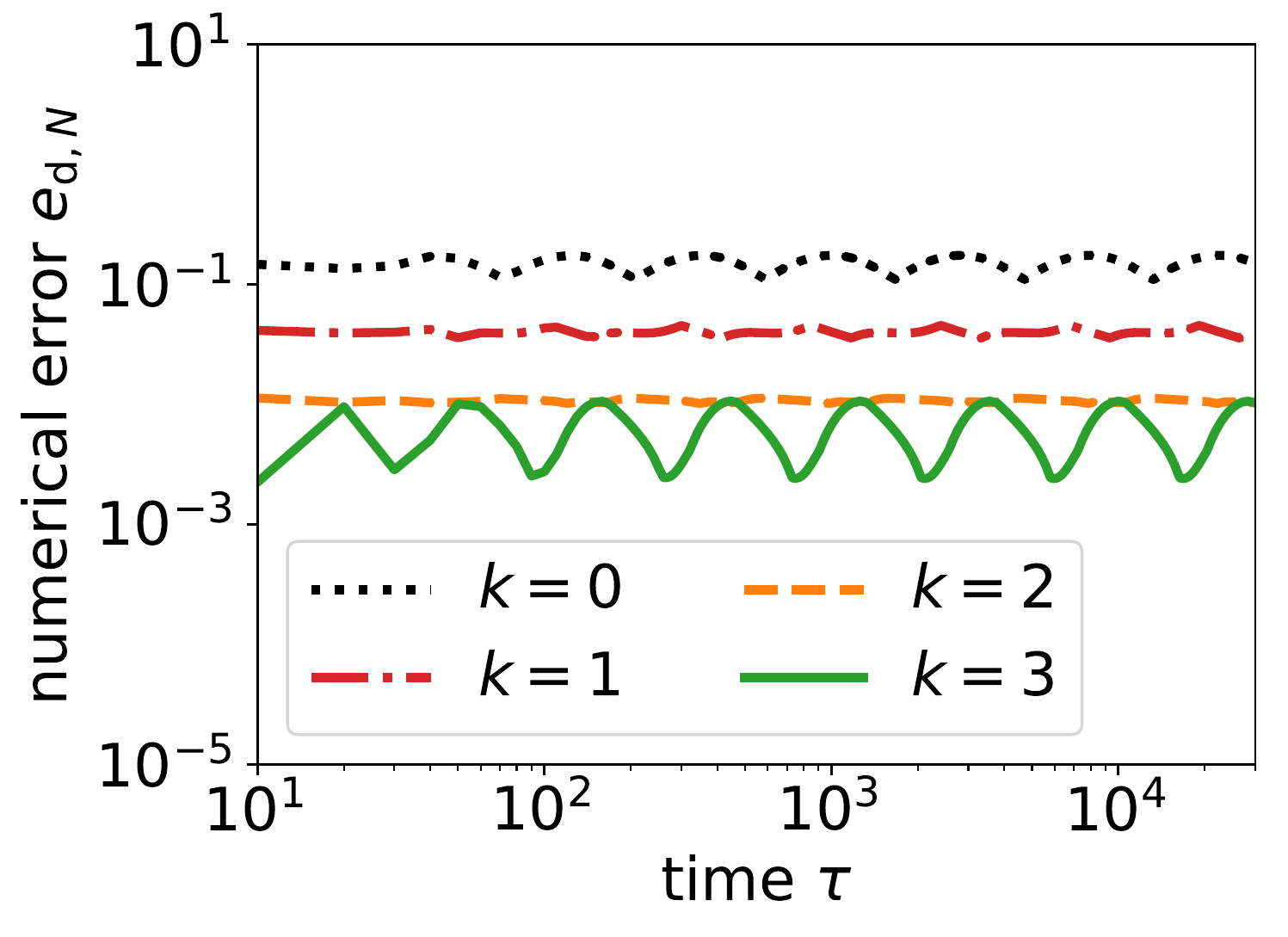}
\caption{Test case, constant kernel: numerical errors $e_{c,N}$ with the $L^1$ continuous norm, $e_{d,N}$ with the discrete $L^1$ norm. All these errors are calculated for $N=20$. Errors remain bounded at large times.}
\label{fig:kconst_err_L1cont_L1dis}
\end{figure}

Time evolution of the numerical errors $e_{c,N}$ and $e_{d,N}$ are shown in Fig.~\ref{fig:kconst_err_L1cont_L1dis}. The results are shown for $N=20$ bins for $k = 0,1,2,3$ at time $\tau = 30000$, when particles of large masses have formed. We verify that $e_{c,N}$ and $e_{d,N}$ remain bounded.

\subsubsection{Computational efficiency}
\label{sec:kconst_computational_performance}
\begin{figure}
\centering
\includegraphics[width=\columnwidth]{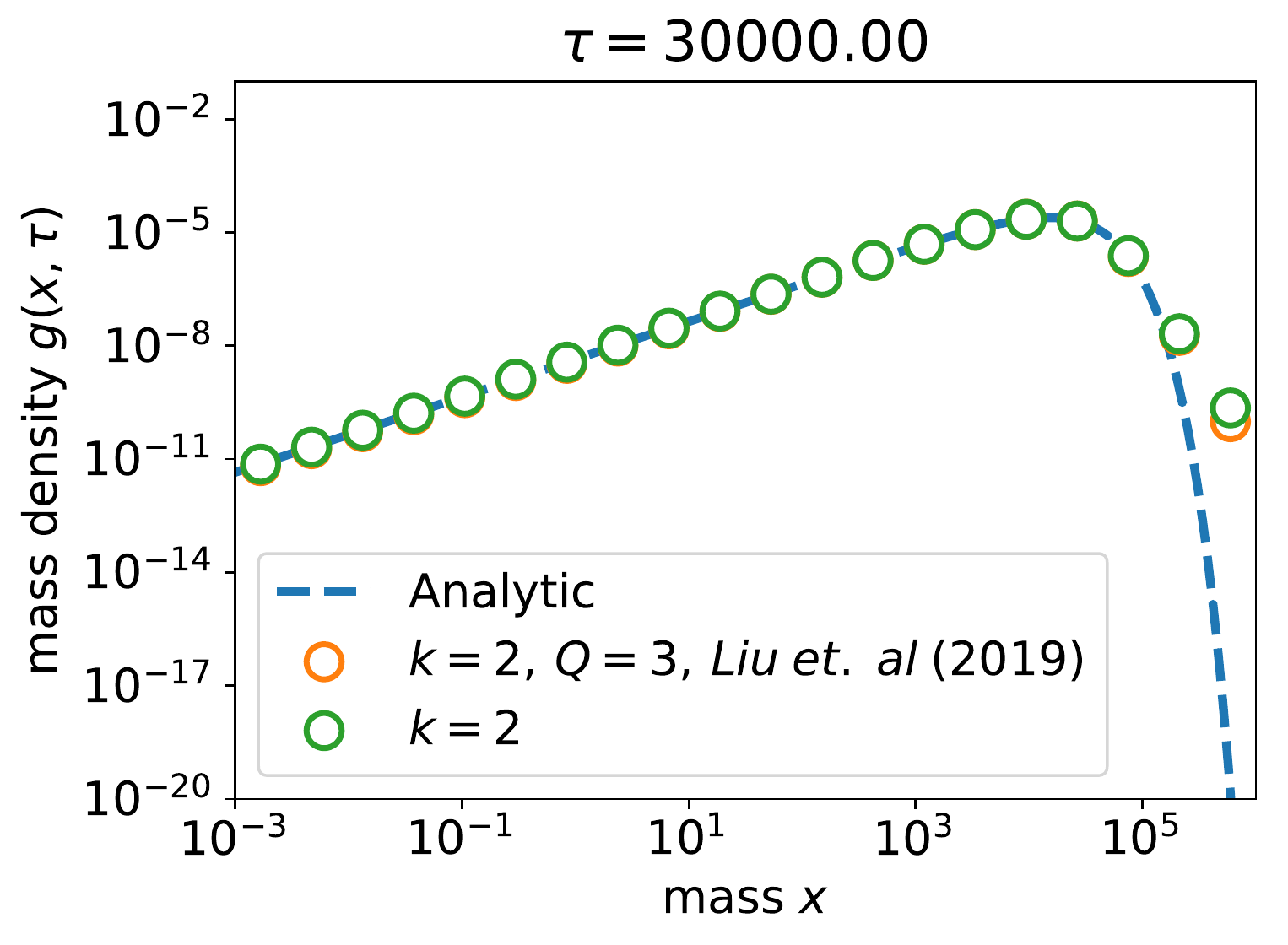}
\caption{Test case, constant kernel: comparison with the scheme of \citet{Liu2019}. Similar accuracies are reached, but being $\sim 4\times$ more effective due to numerical integration.}  
\label{fig:kconst_loglog_xmeanlog_DGvsDGGQ}
\end{figure}

Fig.~\ref{fig:kconst_loglog_xmeanlog_DGvsDGGQ} shows that similar accuracies are obtained with this scheme and the scheme described in \cite{Liu2019}. Computational time is compared on a simulation with $N=20$ bins, $k=2$ and a final time $\tau = 30000$ after $\sim 10^{3}$ timesteps. The computational time for the \citet{Liu2019} scheme is around 16 seconds (real time). The computational time for this scheme is around 4 seconds (real time). An improvement of factor 4 is therefore achieved for the computational time by estimating integrals analytically.

\subsection{Additive kernel}
\label{sec:kadd_tests}

\subsubsection{Positivity and mass conservation}
\begin{figure*}
\centering
\includegraphics[width=0.8\textwidth]{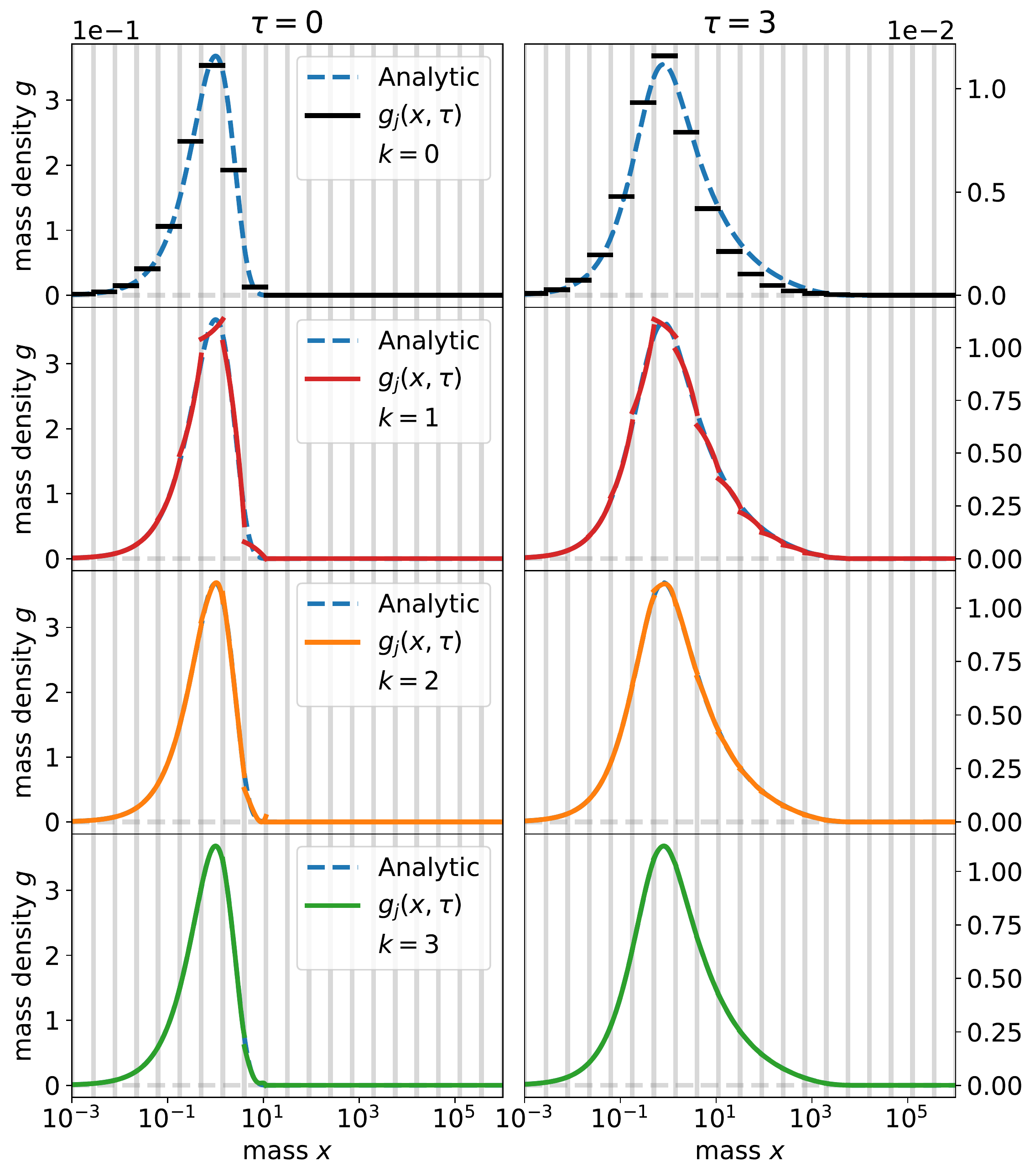}
\caption{Test case, additive kernel: the numerical solution $g_j(x,\tau)$ is plotted for $N=20$ bins and $k = 0,1,2,3$ from $\tau=0$ to $\tau=3$, and compared to the analytic solution $g(x,\tau)$. Vertical grey lines delimit the bins. The accuracy improves for larger values of $k$. Order $3$ approximates the bump where the major part of the mass is concentrated with accuracy of order $\sim 0.1 \%$.
}
\label{fig:kadd_linlog}
\end{figure*}

\begin{figure}
\centering
\includegraphics[width=0.95\columnwidth]{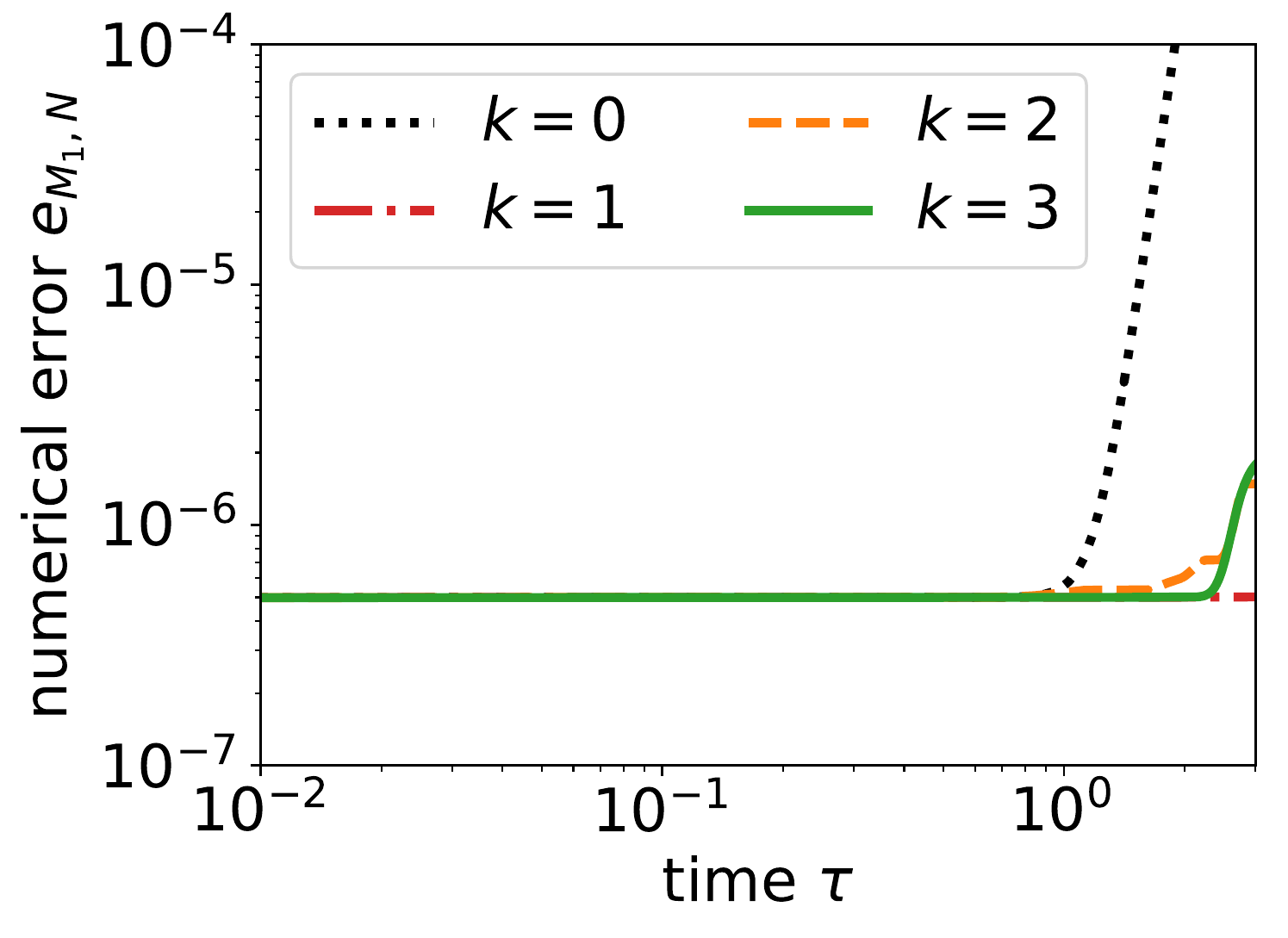}
\caption{Test case, additive kernel: evolution of the numerical absolute error $e_{M_1,N}$ on the moment $M_{1,N}$ for $N=20$ bins. The divergence at long times is explained by accumulation of errors due to numerical diffusion for orders $k = 0$, $k = 2$ and $k = 3$. Total mass is conserved at machine precision until $\tau=1$.
}
\label{fig:mass_cons_kadd}
\end{figure}

Fig.~\ref{fig:kadd_linlog} shows numerical solutions obtained for $N=20$ bins and $k = 0,1,2,3$ at time $\tau=3$. The numerical solutions remains positive as grains grow.  Fig.~\ref{fig:mass_cons_kadd} shows the evolution of the numerical absolute error $e_{M_1,N}$ on the first moment $M_{1,N}$. The total mass remains conserved to machine precision until $\tau=1$.

\subsubsection{Accuracy of the numerical solution}
\label{sec:kadd_accuracy}
\begin{figure}
   \centering
   \includegraphics[width=\columnwidth]{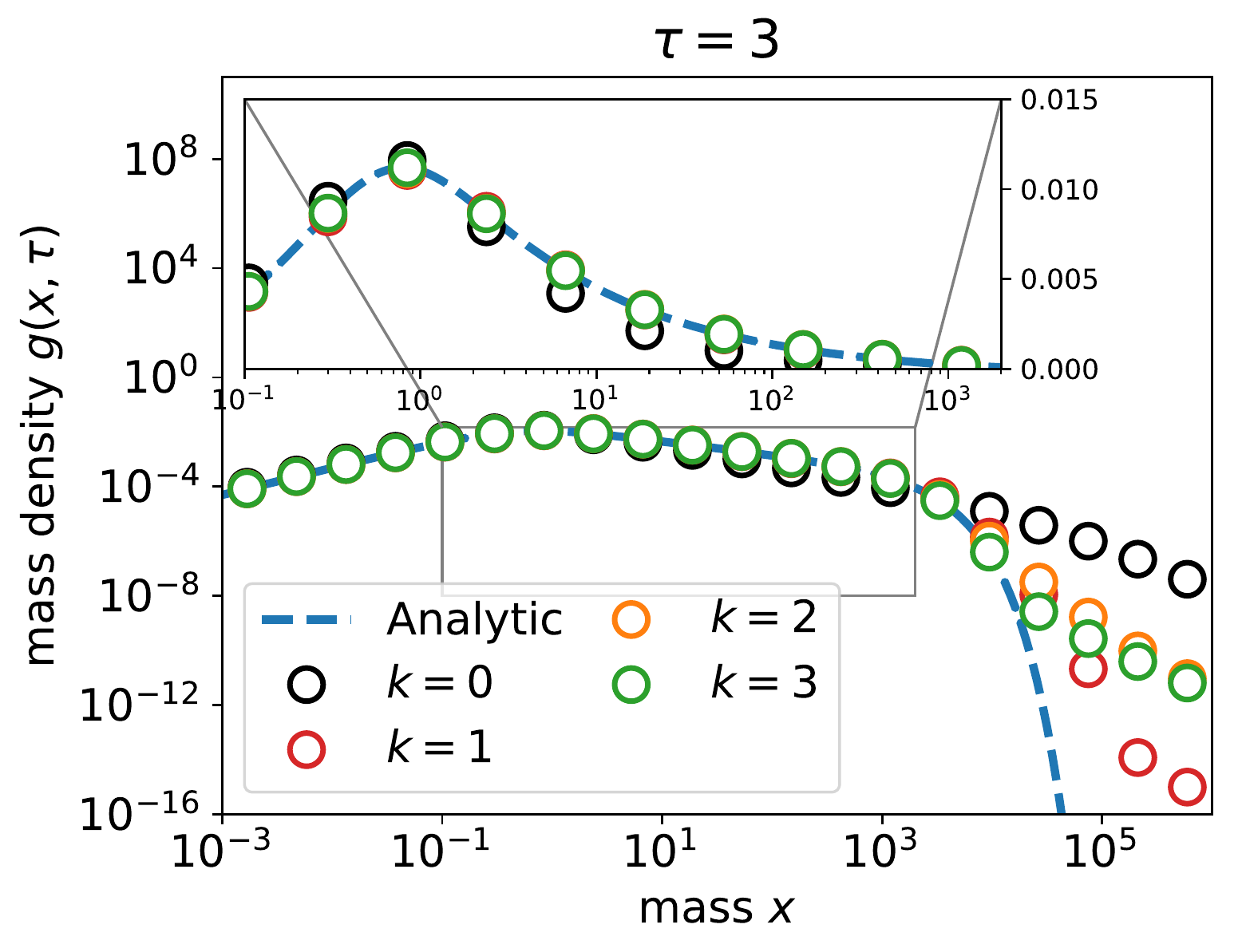}
   \caption{
Test case, additive kernel: At the location of the maximum, orders $k = 1,2,3$ achieve an absolute error of $\sim 0.1 - 1\%$, to be compared with $10\%$ obtained with $k = 0$. Accuracy in the exponential tail is improved by a factor $10000$ by $k = 3 $ compared to $k = 0$.
   }  
   \label{fig:kadd_loglog_xmeanlog}
\end{figure}

Fig.~\ref{fig:kadd_loglog_xmeanlog} shows numerical solutions obtained at $\tau=3$ on a logarithmic scale. Fig.~\ref{fig:kadd_loglog_xmeanlog} reveals a strong numerical diffusion for order $0$. Numerical errors are indeed integrated and diffused extremely efficiently towards large masses by the additive kernel. In this case, the mass density for  large-masses particles is over-estimated by several orders of magnitude. High-order schemes reduce this numerical diffusion as expected. Fig.~\ref{fig:kadd_loglog_xmeanlog} shows that around the maximum, schemes of order $k = 1,2,3$ provide errors of order $\sim 0.1 - 1\%$ when $k = 0$ generates errors of order $\sim 10\%$. Numerical diffusion is reduced in the exponential tail as the order of the scheme increases, up to reaching a gain of a factor $\sim 10000$ with order $3$ compared to order $0$.

\subsubsection{Convergence analysis}
\label{sec:kadd_convergence}
\begin{figure}
\centering
\includegraphics[width=\columnwidth]{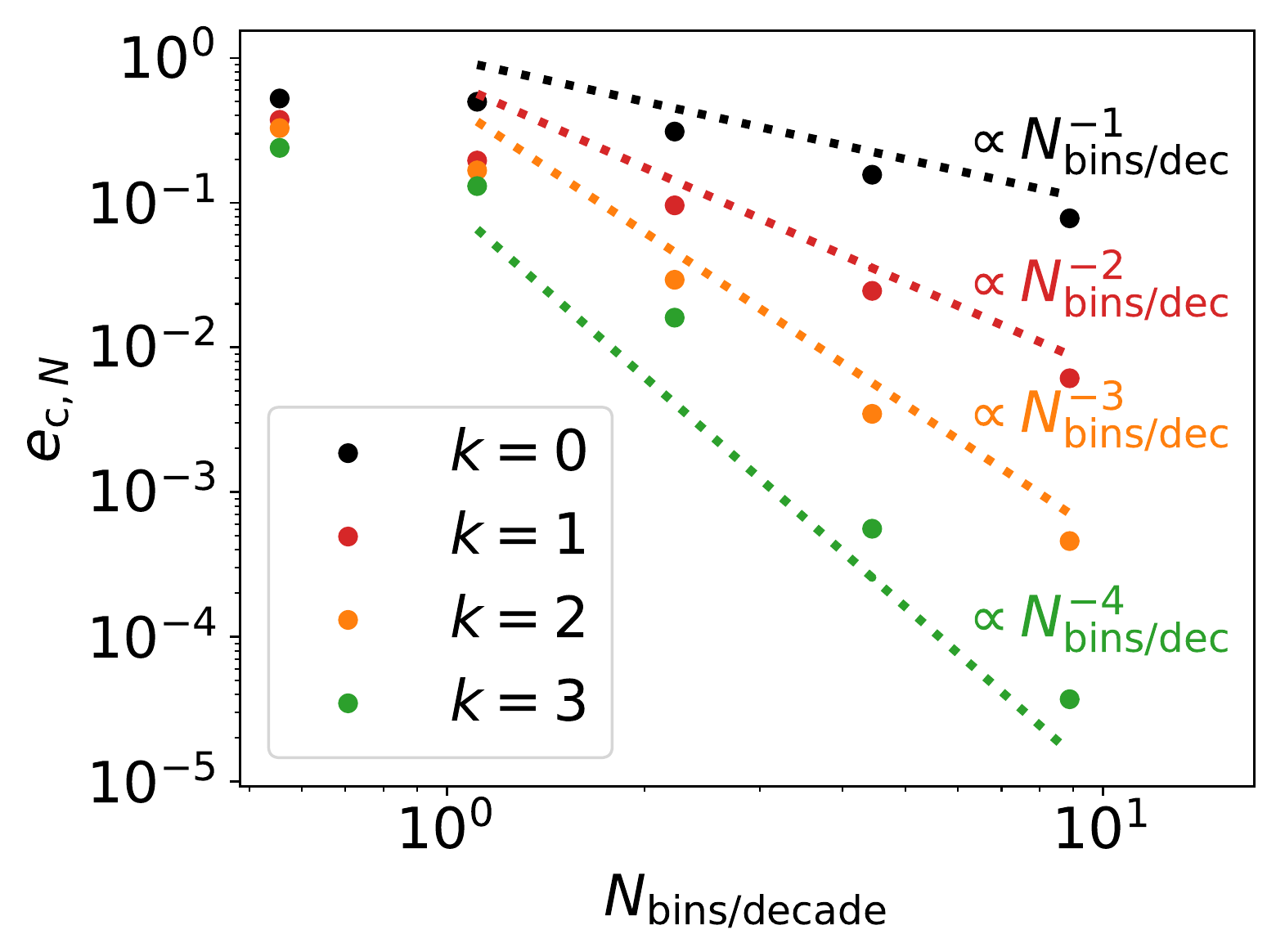}
\includegraphics[width=\columnwidth]{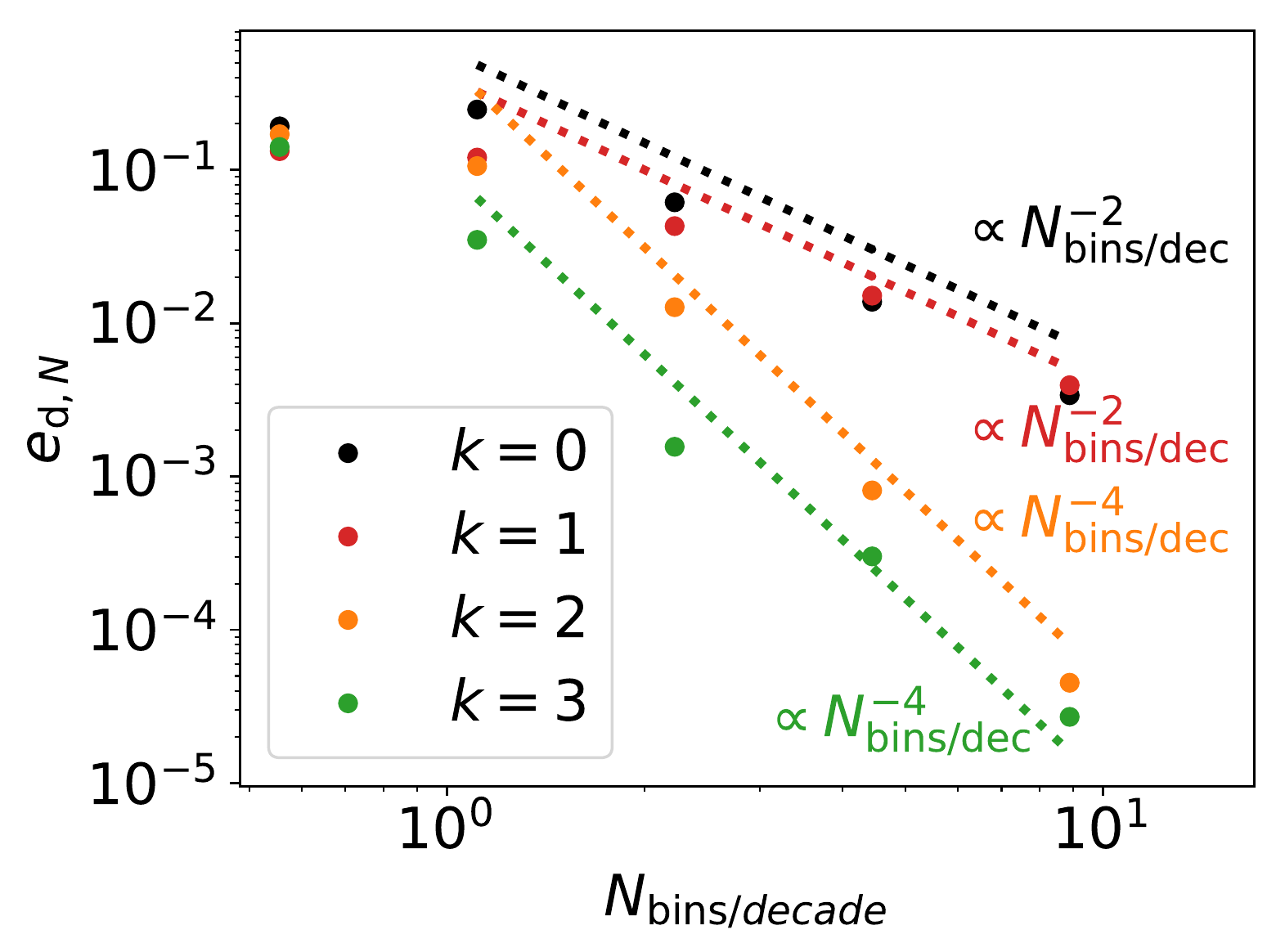}
\caption{Test case, additive kernel: similar to Fig.~\ref{fig:kconst_errL1_convergence}. The DG scheme achieves on $e_{d,N}$ an accuracy of order $0.1\%$ with more than $10$ bins/decade for $k=0,1$, with $\sim 5$ bins/decade for $k = 2,3$. An accuracy of order $1\%$ is achieved with $\sim 9$ bins/decade for $k = 0,1$, with $\sim 5$ bins/decade for $k = 2$ and with $\sim 2$ bins/decade for $k = 3$.}
\label{fig:kadd_errL1_convergence}
\end{figure}

Numerical errors are shown on Fig.~\ref{fig:kadd_errL1_convergence} at $\tau=0.01$. Accuracy of order $\sim 0.1\%$ on $e_{d,N}$ errors are achieved with more than $10$ bins/decade for order $0$ and $1$, with $\sim 9$ bins/decade for orders $2$ and $3$. Accuracy of order $\sim 1\%$ is achieved with $\sim 9$ bins/decade for orders $0$ and $1$, with $\sim 5$ bins/decade for order $2$ and $\sim 2$ bins/decade for order $3$. 

\subsubsection{Stability in time}
\label{sec:kadd_stability_time}
\begin{figure}
\centering
\includegraphics[width=\columnwidth]{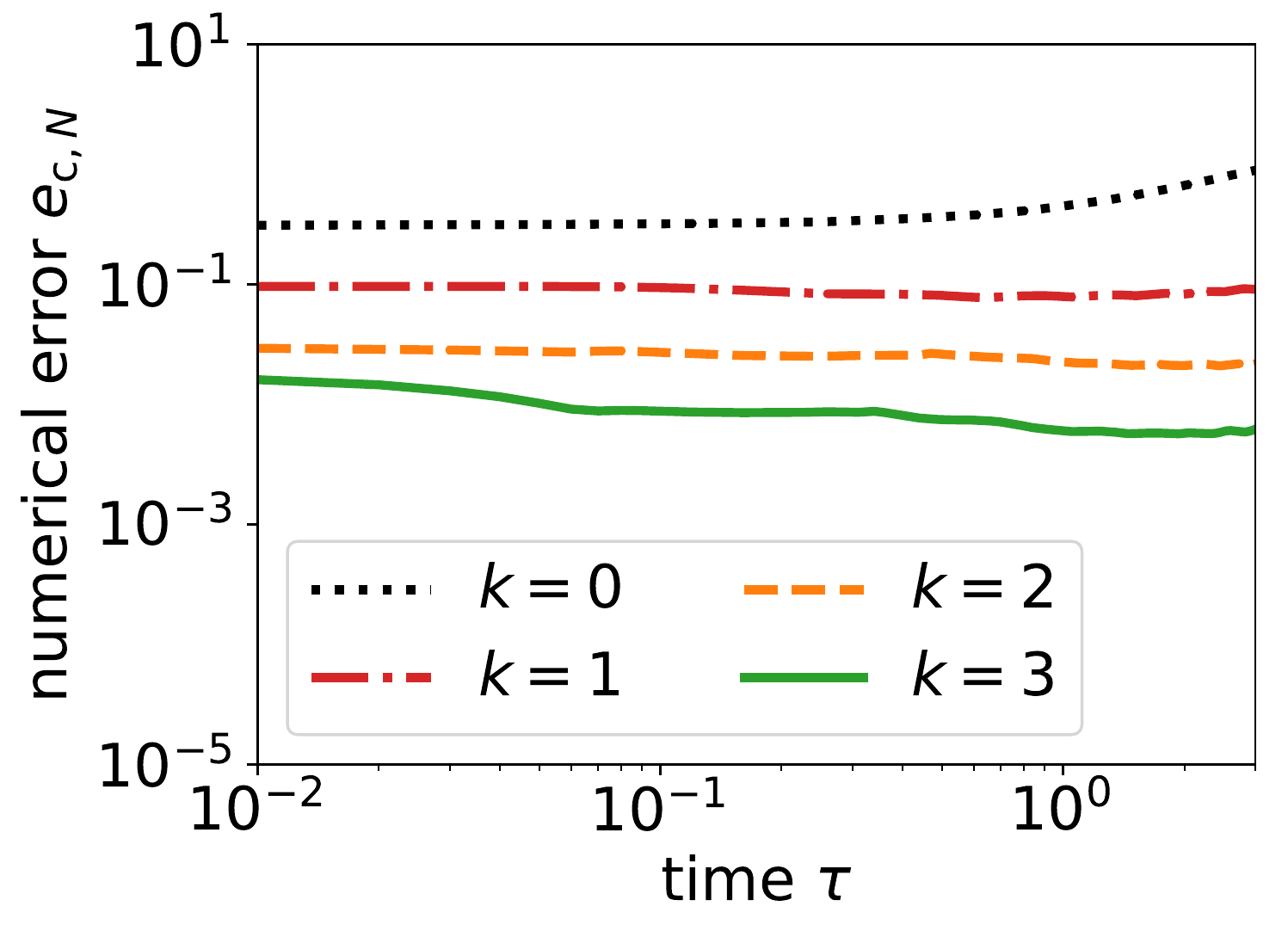}
\includegraphics[width=\columnwidth]{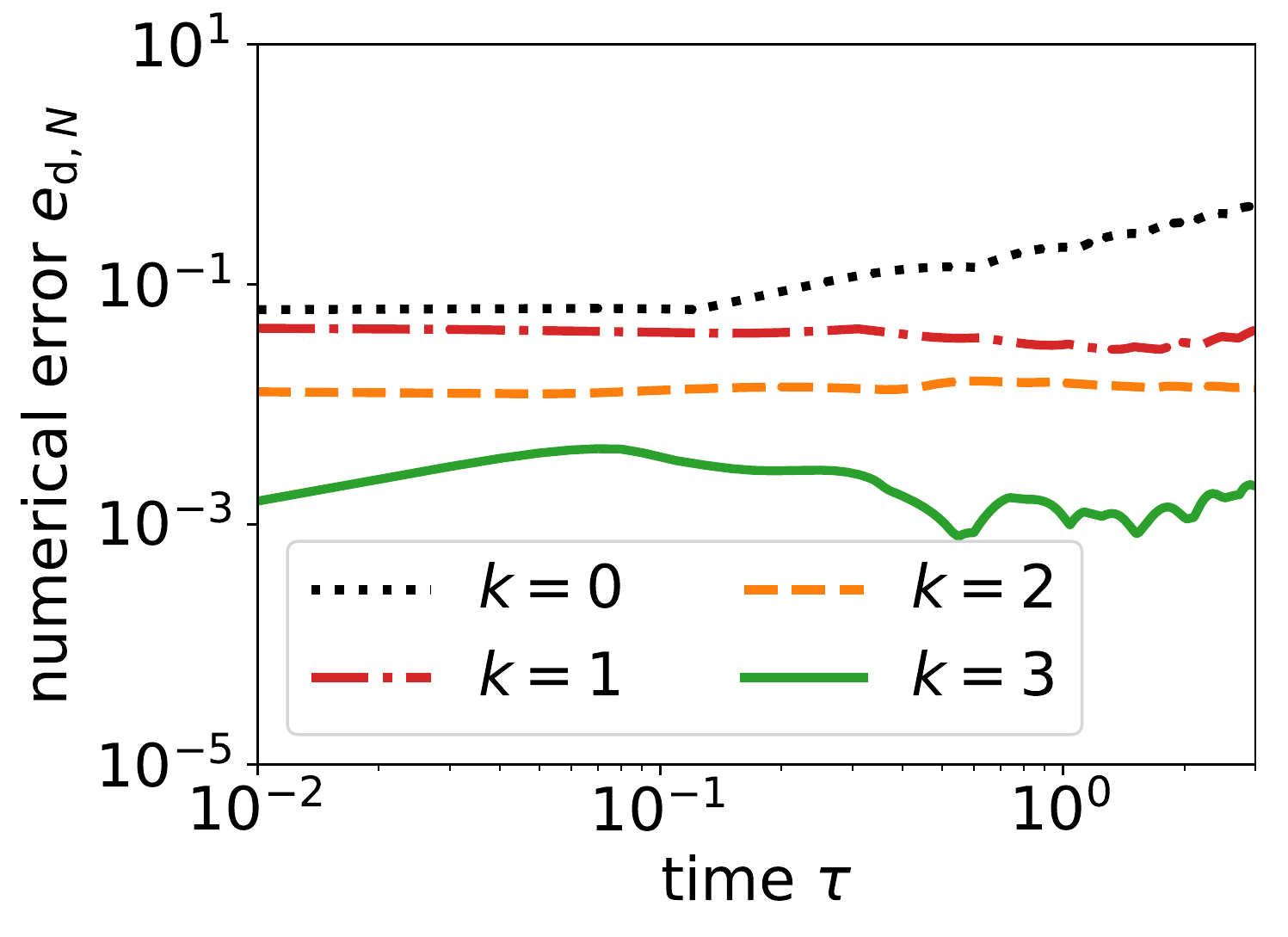}
\caption{Test case, additive kernel: numerical errors $e_{c,N}$ with the $L^1$ continuous norm, $e_{d,N}$ with the discrete $L^1$ norm. All these errors are calculated for $N=20$. Errors remain bounded at large times for orders $k = 1,2,3$.}
\label{fig:kadd_err_L1cont_L1dis}
\end{figure}

Evolution of the numerical errors $e_{c,N}$ and $e_{d,N}$ are shown in Fig.~\ref{fig:kadd_err_L1cont_L1dis}. The results are shown for $N=20$ bins for $k = 0,1,2,3$ at $\tau = 3$, when particles with large masses have formed. At order $0$, $e_{c,N}$ (resp. $e_{d,N}$) increases significantly after $\tau \approx 5\cdot 10^{-1}$ (resp. $\tau \approx 10^{-1}$). On the contrary, $e_{c,N}$ and $e_{d,N}$ remain bounded for longer times at orders $1$, $2$ and $3$. 

\subsubsection{Computational efficiency}
\label{sec:kadd_computational_performance}
\begin{figure}
\centering
\includegraphics[width=\columnwidth]{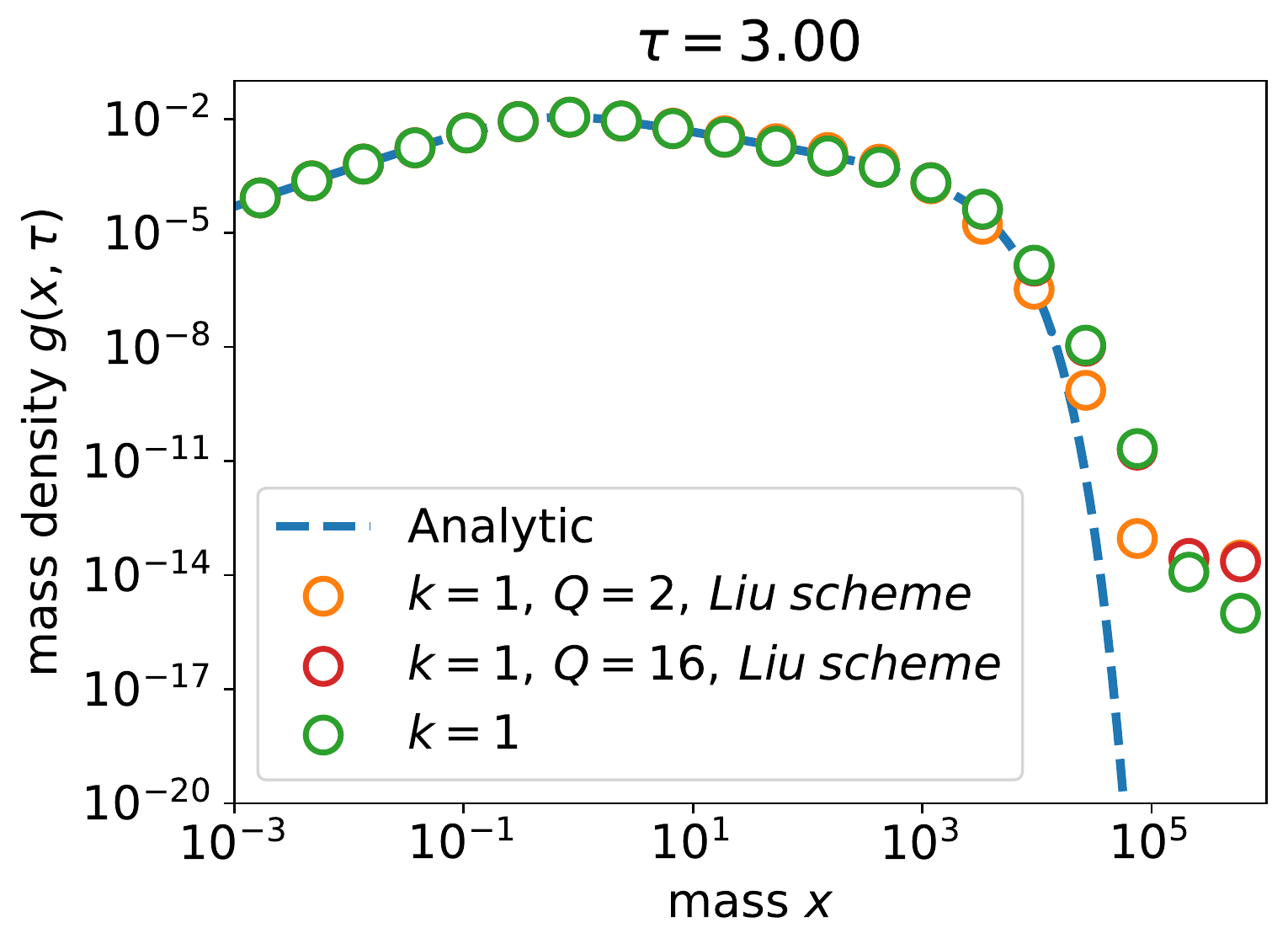}
\caption{Test case, additive kernel: comparison with the scheme of \citet{Liu2019}. Unexpected accuracy occurs for integral estimates with $Q =2$ Gauss points due to serendipitous error compensations. Our algorithm is $\sim 3\times$ more effective due to analytical integration compared to the Lui scheme with $Q=2$. 
}  
\label{fig:kadd_loglog_xmeanlog_DGvsDGGQ}
\end{figure}

Computational time is compared to \citet{Liu2019} on a simulation with $N=20$ bins, $k=1$ and a final time $\tau=3$. Fig.~\ref{fig:kconst_loglog_xmeanlog_DGvsDGGQ} shows similar accuracy for both schemes. The computational time for the \citet{Liu2019} scheme is around 3 seconds (real time) for a number of Gauss quadrature points $Q=2$. The computational time for this scheme is 1 second, providing an improvement by a factor 3. Fig.~\ref{fig:kadd_loglog_xmeanlog_DGvsDGGQ} also shows that for the additive kernel, the Liu scheme with $Q=2$ is counter-intuitively more accurate than for $Q=16$ and the DG scheme. This result can be explained by a serendipitous compensation of errors when approximating the integrals with a Gauss quadrature of low order.

\subsection{Multiplicative kernel}
\label{sec:kmul_tests}

\subsubsection{Positivity and mass conservation}
\begin{figure*}
\centering
\includegraphics[width=0.8\textwidth]{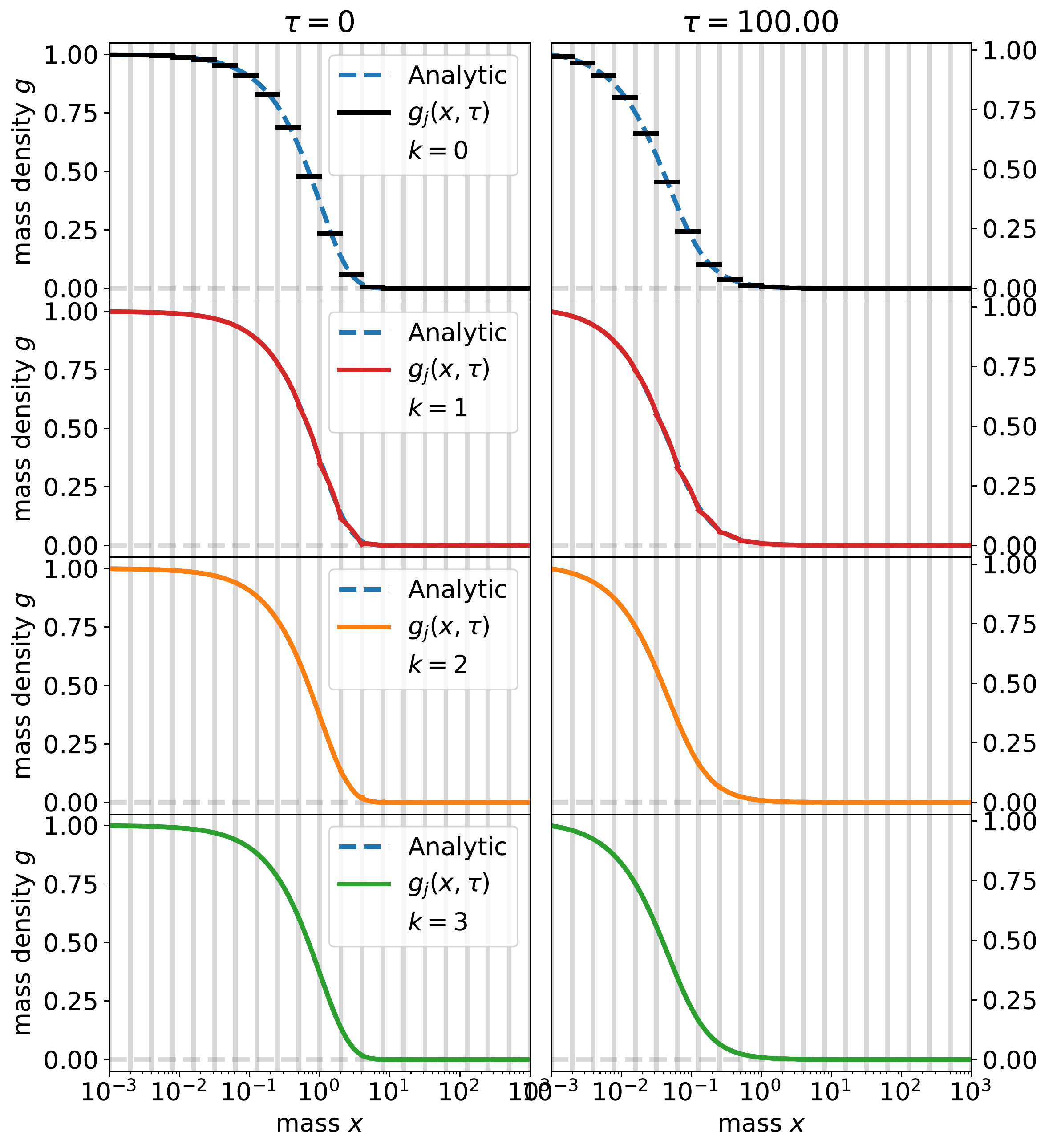}
\caption{Test case, multiplicative kernel: numerical solution $g_j(x,\tau)$ is plotted for $N=20$ bins for $k = 0,1,2,3$ from $\tau=0$ to $\tau=100$ and compared to the analytic solution $g(x,\tau)$. Vertical grey lines delimit the bins. Accuracy of order $\sim 0.1 \%$ is achieved at all orders.
}
\label{fig:kmul_linlog}
\end{figure*}

\begin{figure}
\centering
\includegraphics[width=\columnwidth]{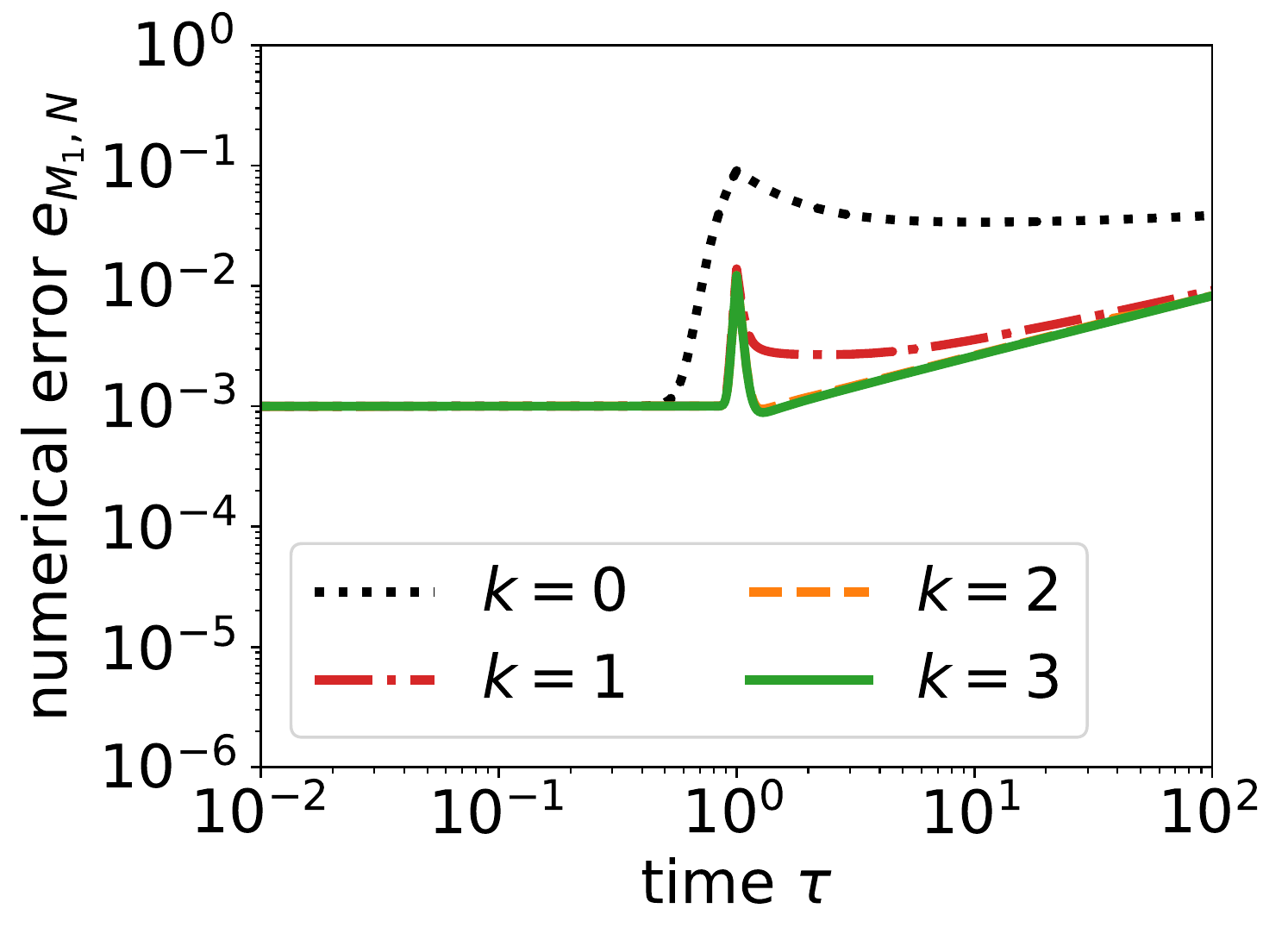}
\includegraphics[width=\columnwidth]{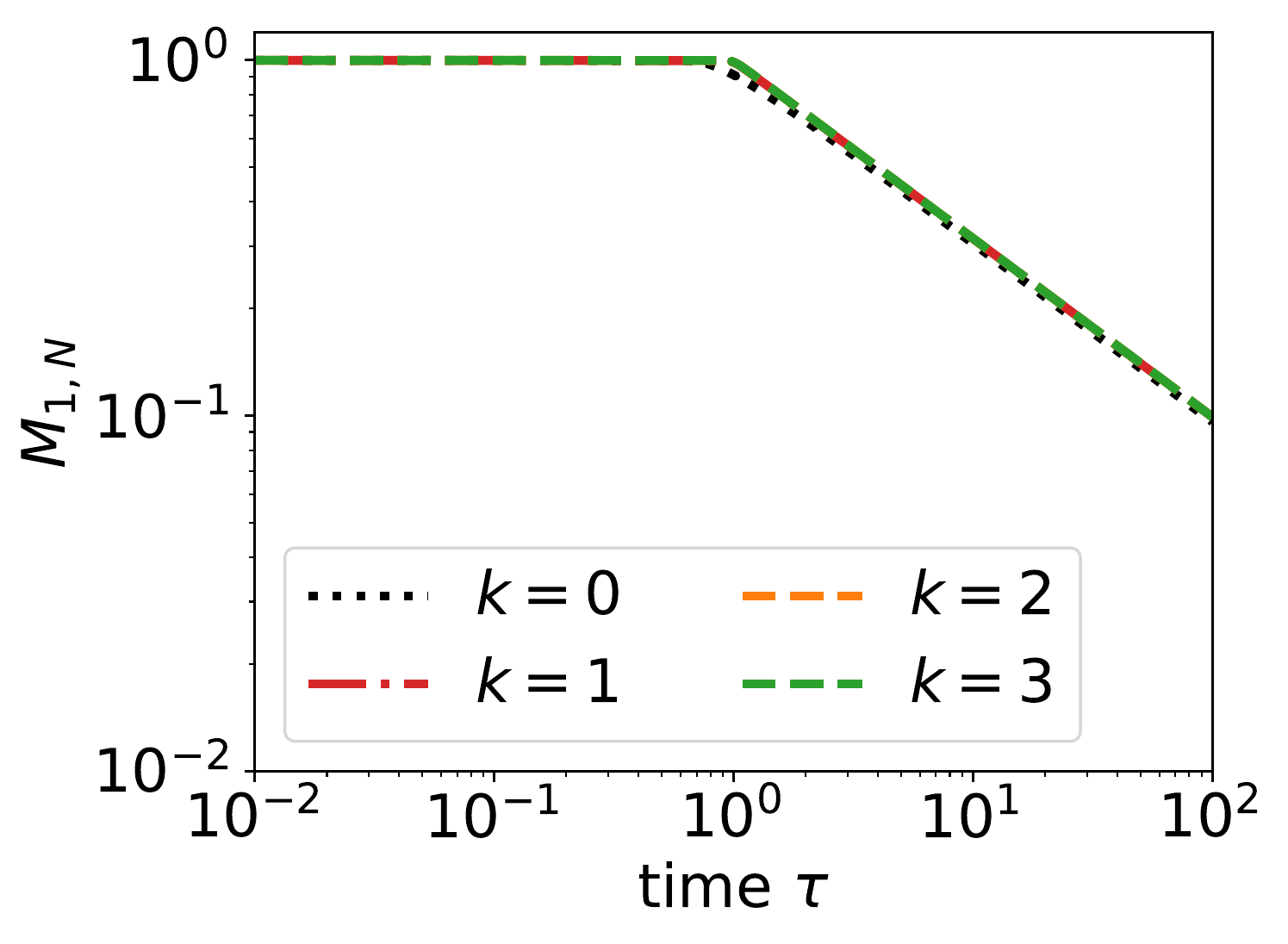}
\caption{
Test case, multiplicative kernel: evolution of the numerical absolute error $e_{M_1,N}$ on the moment $M_{1,N}$ for $N=20$ bins. Mass is conserved anymore when gelation occurs at $\tau = 1$.
}
\label{fig:mass_cons_kmul}
\end{figure}

Fig.~\ref{fig:kmul_linlog} shows the numerical solutions obtained for $N=20$ bins and $k = 0,1,2,3$ after $\tau=100$. The numerical solutions remain positive as grain grow. Fig.~\ref{fig:mass_cons_kmul} shows the evolution of $e_{M_1,N}$. Total mass remains conserved to machine precision until $\tau <1$. At $\tau=1$, gelation occurs, particles with infinite mass are formed \citep{McLeod1964,Ernst1984,FL2004} and total mass is no longer conserved anymore.

\subsubsection{Accuracy of the numerical solution}
\label{sec:kmul_accuracy}
\begin{figure}
\centering
\includegraphics[width=\columnwidth]{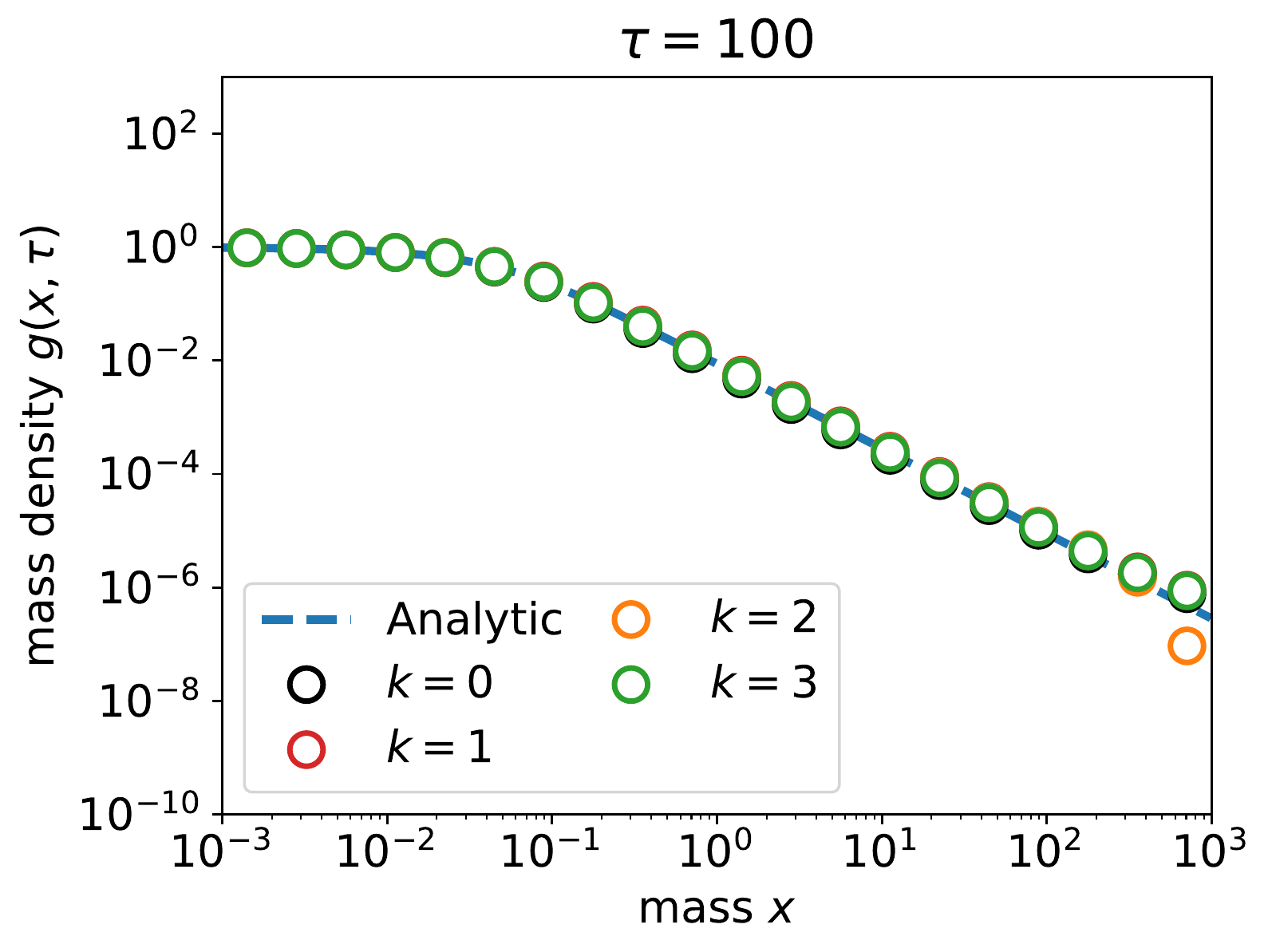}
\caption{Test case, multiplicative kernel: Accuracy of order $\sim 0.1 \%$ is achieved at any order. Growth is so efficient than is overtakes numerical diffusion.
}  
\label{fig:kmul_loglog_xmeanlog}
\end{figure}

Fig.~\ref{fig:kmul_loglog_xmeanlog} shows the numerical solution for the multiplicative kernel at $\tau= 100$. Accuracy of order $\sim 0.1 \%$ is obtained at all orders, even $k = 0$. Physically, growth is effective enough for advection in the mass space to be more efficient than numerical diffusion. 

\subsubsection{Convergence analysis}
\label{sec:kmul_convergence}
\begin{figure}
\centering
\includegraphics[width=\columnwidth]{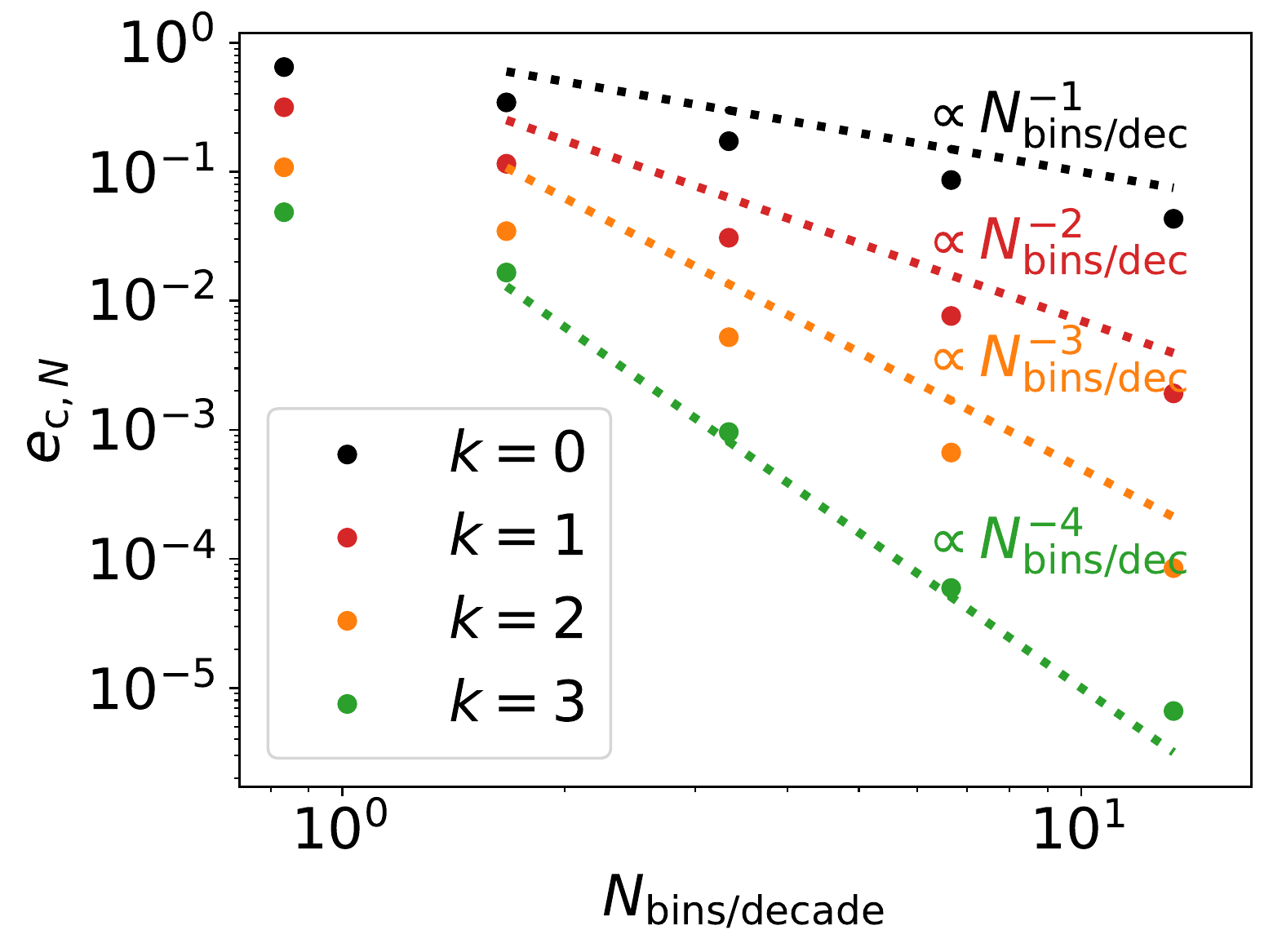}
\includegraphics[width=\columnwidth]{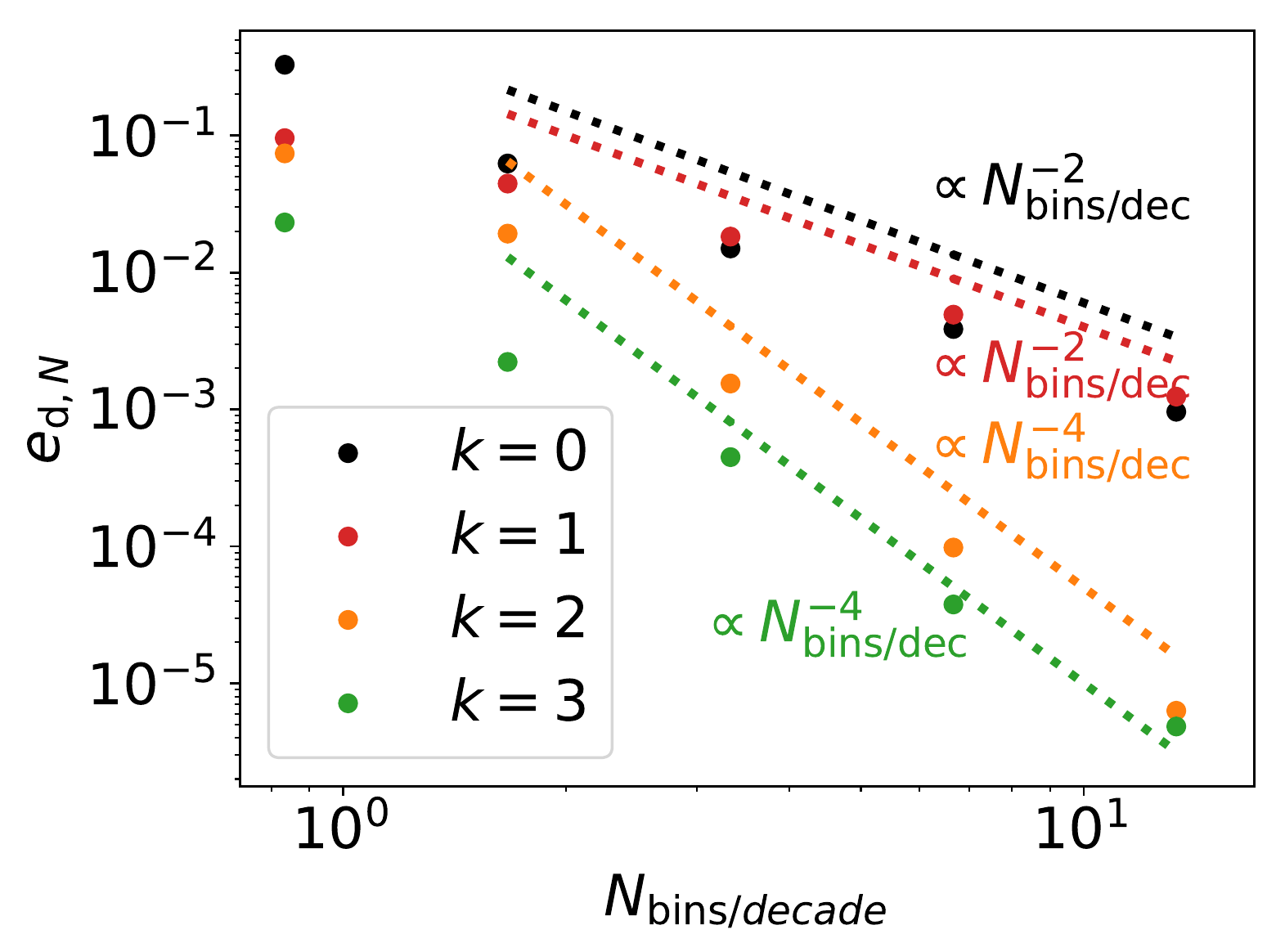}
\caption{Test case, multiplicative kernel: the continuous $L^1$ error $e_{c,N}$ and the discrete $L^1$ error $e_{d,N}$ are plotted as functions of the number of bins per decade. With $e_{c,N}$, the experimental order of convergence is EOC = $k+1$. With $e_{d,N}$, EOC = $k+1$ for polynomials of odd orders and EOC = $k+2$ for polynomials of even orders. The DG scheme achieves on $e_{d,N}$ an accuracy of $0.1\%$ with $\sim 15$ bins/decade for $k = 0,1$, with $\sim 7$ bins/decade for $k = 2$ and with $\sim 4$ bins/decade for $k = 3$ . An accuracy of $1\%$ is achieved with $\sim 7$ bins/decade for $k = 0,1$, with $\sim 2$ bins/decade for $k = 2$ and with $\sim 1$ bins/decade for $k = 3$.}
\label{fig:kmul_errL1_convergence}
\end{figure}

Numerical errors are shown on Fig.~\ref{fig:kmul_errL1_convergence} at $\tau=0.01$. Accuracy of order $\sim 0.1\%$ on $e_{d,N}$ errors are achieved with $\sim 15$ bins/decade for orders $0$ and $1$, with $\sim 7$ bins/decade for order $2$, and with $\sim 4$ bins/decade for order $3$. Accuracy of order $\sim 1\%$ is achieved with $\sim 7$ bins/decade for orders $0$ and $1$, with $\sim 2$ bins/decade for order $2$, and with $\sim 1$ bins/decade for order $3$. 

\subsubsection{Stability in time}
\label{sec:kmul_stability_time}
\begin{figure}
\centering
\includegraphics[width=\columnwidth]{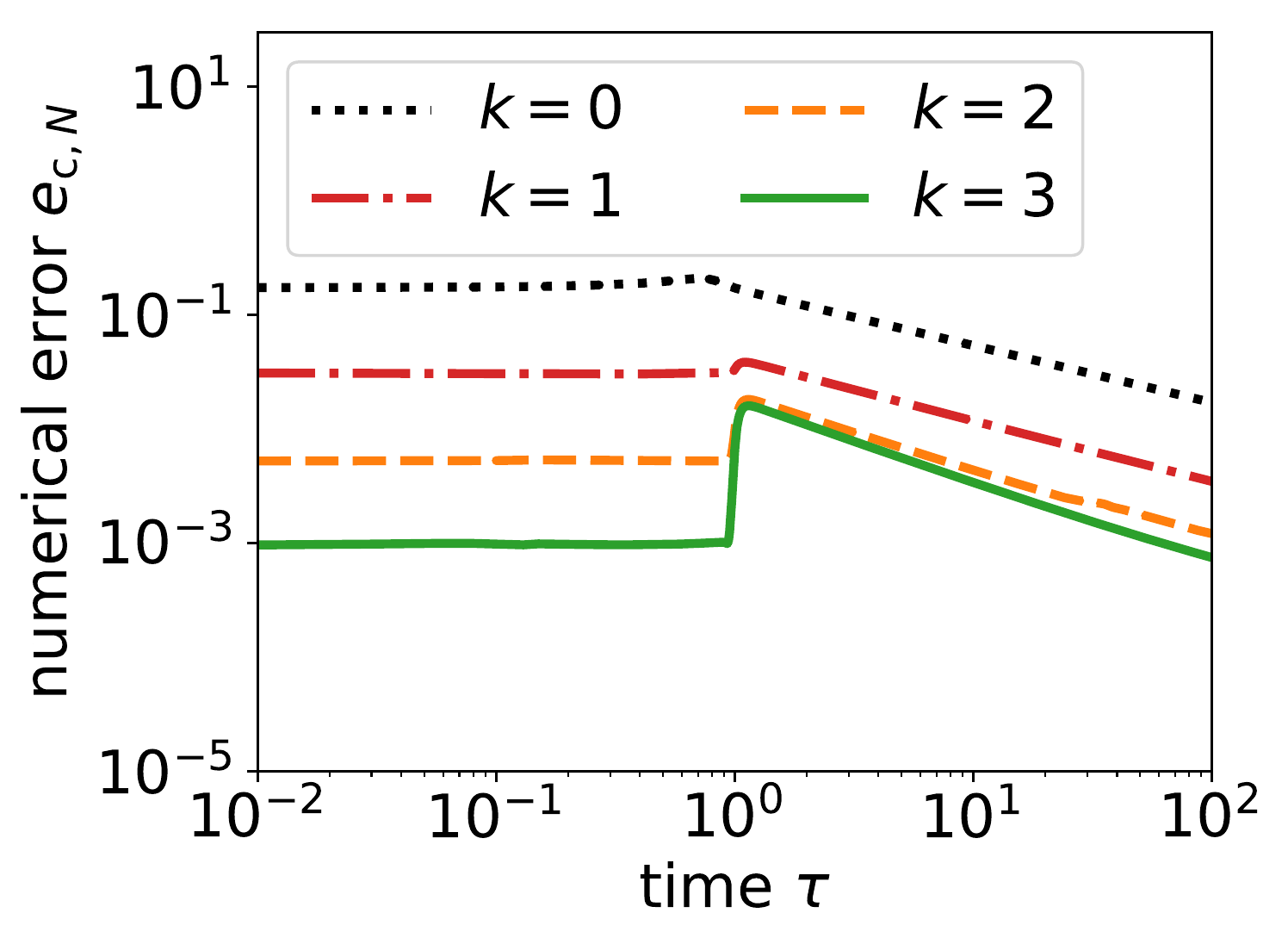}
\includegraphics[width=\columnwidth]{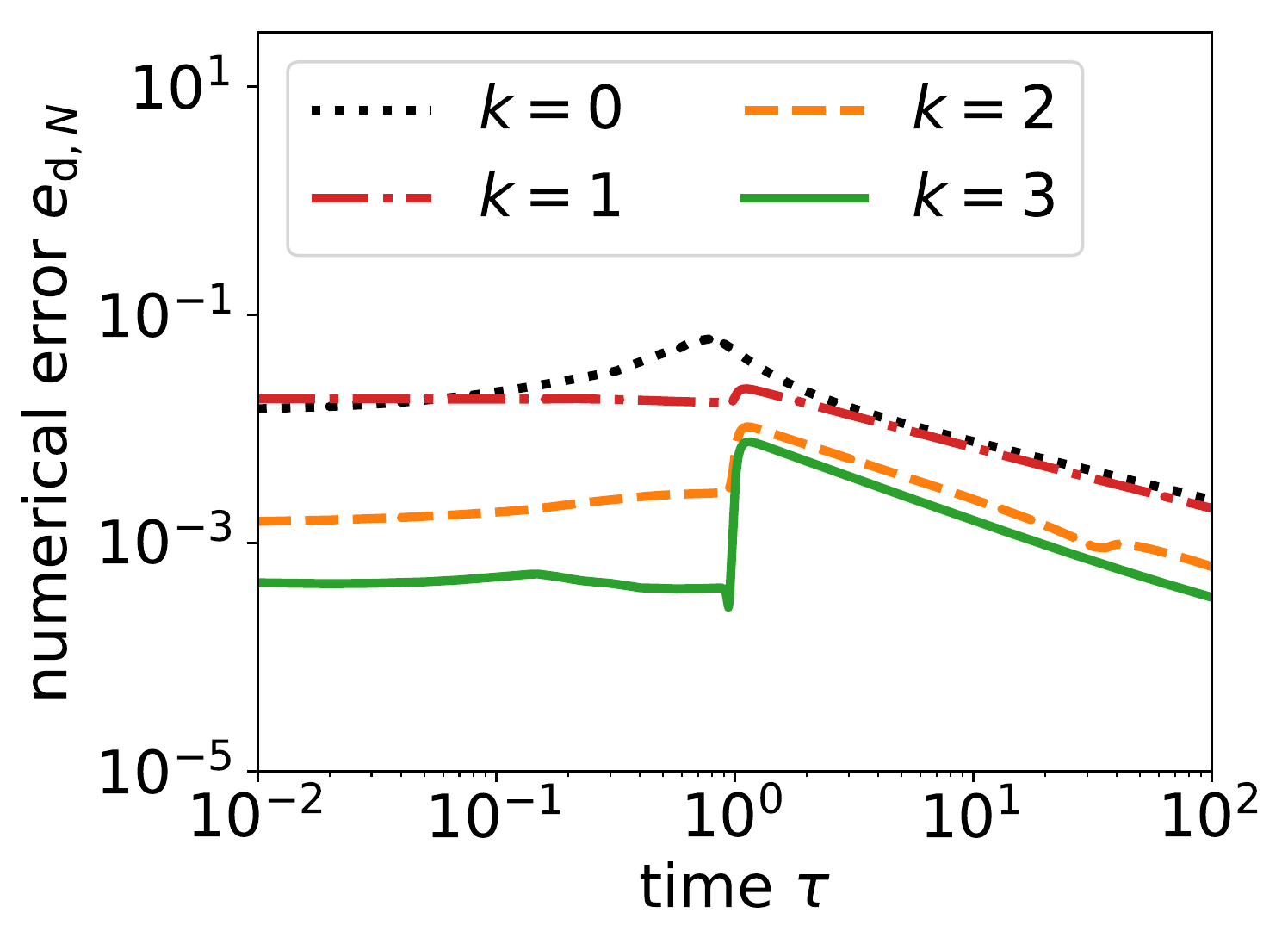}
\caption{Test case, multiplicative kernel: numerical errors $e_{c,N}$ with the $L^1$ continuous norm, $e_{d,N}$ with the discrete $L^1$ norm. All these errors are calculated for $N=20$. Errors remain bounded at large times.
}
\label{fig:kmul_err_L1cont_L1dis}
\end{figure}

The evolution of the numerical errors $e_{c,N}$ and $e_{d,N}$ are shown in Fig.~\ref{fig:kmul_err_L1cont_L1dis}. The results are shown for $N=20$ bins fo $k = 0,1,2,3$ at time $\tau = 100$, when particles with large masses have formed. We observe that $e_{c,N}$ and $e_{d,N}$ remain bounded, even after the occurence of gelation at $\tau=1$.

\subsubsection{Computational efficiency}
\label{sec:kmul_computational_performance}
\begin{figure}
\centering
\includegraphics[width=\columnwidth]{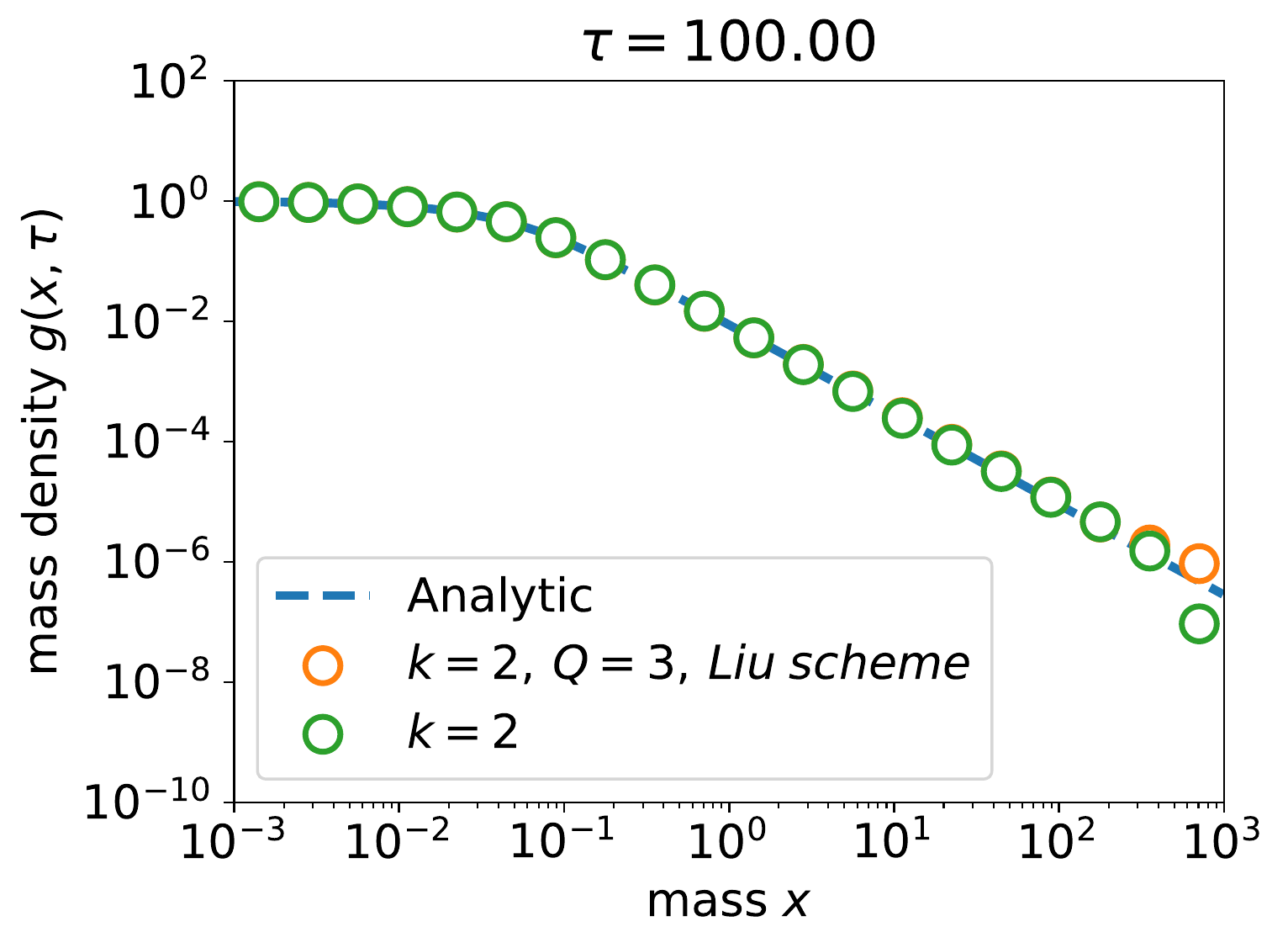}
\caption{Test case, multiplicative kernel: comparison between the numerical solutions provided by this scheme and the scheme of \citet{Liu2019}. Similar accuracies are reached, but being $\sim 5\times$ more effective due to analytical integration.}
\label{fig:kmul_loglog_xmeanlog_DGvsDGGQ}
\end{figure}

Fig.~\ref{fig:kmul_loglog_xmeanlog_DGvsDGGQ} shows similar accuracies for the \citet{Liu2019} scheme and our implementation. With $k = 2$, the computational time for the \citet{Liu2019} scheme is around 8 minutes for a number of Gauss quadrature points $Q=3$. The computational time is for this scheme 1 minute and 40 seconds, providing an improvement by a factor 5.

\section{Discussion}
\label{sec:discussions}

The Discontinuous Galerkin scheme presented in Sect.~\ref{sec:dg} involves polynomials of high-order, implying issues with differences of large real numbers. Order $k = 3$ appears as a maximum limit for the order of the scheme in its current form in practice. So far, the ratio betwen $10^{6}$ coagulation time-steps and one hydrodynamical time-step with \texttt{PHANTOM} using $10^{6}$ SPH particles is of order $\sim 10-100$. We are confident the we can reach a one-to-one ratio by i) taking advantage of more ingenious time-stepping (e.g. \citealt{Carrillo2004,Goudon2013}, ii)  adopt a more relevant choice for the basis (e.g. \citealt{Soong1974}) and iii) use GPU parallelisation, since calls  to the coagulation solver by the hydrodynamical code are independent. These strategies to further gain accuracy and computational efficiency will be tested in a next future.

The most relevant kernel for astrophysics is the ballistic kernel (Sect.~\ref{sec:kernels}). Large-scale values of $\Delta v$ are provided by 2D piecewise constant functions from hydrodynamic codes. In discs, the $\Delta v$ function encompasses radial drift, vertical settling and turbulence at large scales. The ballistic kernel splits in three terms
\begin{equation}
  \begin{aligned}
    \mathcal{K}_{b}(u,v) &= \pi (u^{2/3}+2u^{1/3}v^{1/3} + v^{2/3}) \Delta v(u,v) \\
    & = \mathcal{K}_{b,1}(u,v) + \mathcal{K}_{b,2}(u,v) + \mathcal{K}_{b,3}(u,v),
  \end{aligned} 
\end{equation}
$\mathcal{K}_{b,1}(u,v) \equiv \pi u^{2/3} \Delta v(u,v)$, $\mathcal{K}_{b,2}(u,v) \equiv \pi 2u^{1/3}v^{1/3} \Delta v(u,v) $ and $\mathcal{K}_{b,3}(u,v) \equiv \pi v^{2/3} \Delta v(u,v)$. The numerical flux is also split in three terms that are evaluated analytically. Models of differential velocities are also used to model sub-grid small-scale values of $\Delta v$ (Brownian motion, dusty turbulence at small scales). Shall these kernels not be integrable, we will estimate them with an appropriate interpolation.

Moreover, the algorithm presented above solves for the Smoluchowski equation with pure growth. Although fragmentation plays a key role in regulating the number of small grains and preventing the formation of large bodies, it has not being included in the solver yet. The algorithm presented in Sect.~\ref{sec:dg} has been designed to incorporate fragmentation genuinely by adding the extra fragmentation flux \citep{Paul2018}
\begin{equation}
   \begin{aligned}
      & F_{\mathrm{frag}} \left[ g \right] \left( x,\tau \right) \equiv \\
      & \int\limits_0^{\infty} \int\limits_x^{\infty} \int\limits_0^x \frac{w}{yz}b(w,y,z)\mathcal{K}(y,z)g(y,\tau)g(z,\tau)\mathrm{d}w \mathrm{d}y \mathrm{d}z,
   \end{aligned}
\label{eq:frag_cont_cons}
\end{equation}
similarly e.g. to \citet{Birnstiel2010}. The kernel $\mathcal{K}$ provides the fragmentation rate between two particles of masses $x$ and $y$. The function $b$ is the breakage rate related to the formation of a particle of mass $x$ from particles of mass $y$ and $w$. Known functional forms of the fragmentation kernel should authorise direct analytic integrations, similarly to the derivations performed in Sect.~\ref{sec:fluxes}. For peculiar regimes, fragmentation kernels can alternatively be interpolated. Astrophysical mass distributions are expected to be dominated by large grains. Hence, the CFL condition for fragmentation should be similar to the one for growth \citep{Vericel2020}. If so, numerical integration will be performed explicitly. If not, implicit time-stepping can be implemented in a manageable way since the number of dust bins has been kept minimal with analytic integrations (i.e. linear algebra with $\sim 15\times15$ matrices).
 
Eq.~\ref{eq:smolu_cont} restrains dust interactions to binary collisions between aggregates of spherical shapes. Multiple collisions are not expected to play a critical role in astrophysics, since dust volume densities are extremely low. On the other hand, dust aggregates are  expected to be porous or have fractal structures. In  particular, small bodies that have not been recompacted by collisions are expected to be fluffy. Eq.~\ref{eq:smolu_cont} also reduces probability distributions of velocities to their mean values. This approximation may quench grain growth occurring through rare collisional events, e.g. between large bodies having low relative velocities \citep{Windmark2012,Garaud2013}. Finally, growth is in essence stochastic, but fluctuations of the solution can not be computed with Eq.~\ref{eq:smolu_cont}. This is not critical, those being hardly constrained by observations. Although the solver presented in Sect.~\ref{sec:dg} can not be used directly to treat the additional physical processes described above, the method could be adapted to do so. Lastly, extending Eq.~\ref{eq:smolu_cont} to multiple compositions, without or with change of states has been done in other communities. This comes to the cost of multiplying the number of variables by the number of materials considered. The algorithm presented in Sect.~\ref{sec:dg} is a first step towards reducing the number of dust bins to allow for solving for multiple compositions in 3D. This would have strong implications for planet formation, e.g. by handling snow lines consistently and providing constrains for meteoritic data.

\section{Conclusion}
\label{sec:conclusion}

We have presented an high-order algorithm that solves accurately the coagulation equation with a limited number of dust bins ($\sim 15$). Specifically:
\begin{enumerate}
\item Mass is conserved to machine precision for astrophysical kernels,
\item Positivity is guaranteed by combining an appropriate slope-limiter to a Total Variation Diminishing time-stepping,
\item Creating aggregates of masses larger that the mass of the reservoir is mathematically excluded by a control of the growth flux,
\item Errors of order $0.1 - 1 \%$ are achieved by high-order discretisation in time and space that can be modulated for convergence purpose. They shall not dominate the error budget over hydrodynamics,
\item Combining a low number of bins and analytic integrations allows manageable costs in memory and time,
\item Additional physics should be implementable in a versatile way.
\end{enumerate}
The next step consists of performing 3D hydrodynamical simulations of star and planet formation with accurate dust growth. The design of the algorithm allows to implement additional processes such as fragmentation in a genuine way. This solver encourages the reduction of CO$_{2}$ emissions related to computational astrophysics.

\section*{Acknowledgements}

GL acknowledges funding from the ERC CoG project PODCAST No 864965. This project has received funding from the European Union's Horizon 2020 research and innovation programme under the Marie Sk\l odowska-Curie grant agreement No 823823. This project was partly supported by the IDEXLyon project (contract nANR-16-IDEX-0005) under the auspices University of Lyon. We acknowledge financial support from the national programs (PNP, PNPS, PCMI) of CNRS/INSU, CEA, and CNES, France. We used \textsc{Mathematica} \citep{Mathematica}. We thank L. Tine, E. D\'el\'eage, D. Price, D. Mentiplay, T. Guillet, S. Charnoz, R. Teyssier and the anonymous referee for useful comments and discussions.

\newpage
\section*{Data availability}
\label{data_github}
The data and supplementary material underlying this article are available in the repository "growth" on GitHub at \url{https://github.com/mlombart/growth.git}. Figures can be reproduced following the file \texttt{README.md}. The repository contains data and Python scripts used to generate figures.

\label{lastpage}
\bibliographystyle{mnras}
\bibliography{biblio,biblio_num_smo}

\end{document}